\documentclass[aos]{imsart}

\RequirePackage{amsthm,amsmath,amsfonts,amssymb,bm}
\RequirePackage[numbers]{natbib}
\usepackage{nicefrac, graphicx}
\usepackage{multirow, diagbox, adjustbox}
\usepackage{booktabs, longtable, float, enumitem}
\usepackage{bbm}

\RequirePackage[colorlinks,citecolor=blue,urlcolor=blue]{hyperref}

\startlocaldefs
\theoremstyle{plain}
\newtheorem{theorem}{Theorem}
\newtheorem{lemma}[theorem]{Lemma}
\newtheorem{collorary}[theorem]{Collorary}
\theoremstyle{definition}
\newtheorem{definition}[theorem]{Definition}
\endlocaldefs
\begin{document}
\begin{frontmatter}

\title{A test statistic, $h^*$, for outlier analysis}

\begin{aug}
\author[add1,add2,add3,add4]{\fnms{Johan F.}\snm{Hoorn}}
\ead[label=e1]{@}
\and
\author[add1]{\fnms{Johnny K. W.}\snm{Ho}}
\ead[label=e2]{???@???}

\address[add1]{School of Design, The Hong Kong Polytechnic University, Hung Hom, Hong Kong Special Administrative Region}
\address[add2]{Department of Computing, The Hong Kong Polytechnic University, Hung Hom, Hong Kong Special Administrative Region}
\address[add3]{Research Institute for Quantum Technology, The Hong Kong Polytechnic University, Hung Hom, Hong Kong Special Administrative Region}
\address[add4]{Department of Communication Science, Vrije Universiteit, Amsterdam, The Netherlands}
\end{aug}

\begin{abstract}
Outlier analysis is a critical tool across diverse domains, from clinical decision-making to cyber-security and talent identification. Traditional statistical methods for outlier detection, such as Grubbs' test and Dixon's $Q$, are predicated on the assumption of normality and often fail to account for the meaningfulness of exceptional values within non-normally distributed datasets. In this paper, we introduce the $h^*$ ($h$-star) test statistic, a novel parametric and frequentist approach for the evaluation of global outliers that does not require the assumption of normality. Unlike conventional techniques that focus on the removal of outliers to preserve statistical `integrity,' $h^*$ is designed to assess the distinctiveness of outliers as phenomena worthy of investigation in their own right. The $h^*$ statistic quantifies the extremity of a data point relative to its group, providing a measure of confidence and statistical significance analogous to the role of Student's $t$ in comparing means. We detail the mathematical formulation of $h^*$, including the calculation of confidence intervals and degrees of freedom, and present a table of significance levels. The utility of $h^*$ is demonstrated using empirical data from a study on mood interventions with social robots, highlighting its capacity to discern between stable extraordinary deviations and values that merely appear extreme under conventional criteria. Methodologies of choice of candidate outliers for the test, sample size, Bayesian inference and paired analysis are discussed. A generalisation of the $h^*$ statistic is subsequently proposed, with individual weighting assigned to differences for nuanced contextual description, and a variable exponent for the optimisation of objective inference and the specification of subjective inference. The physical significance of an outlier characterised by the $h^*$-test is then extended to the signature of unique occurrences. Our findings suggest that the $h^*$ test statistic offers a robust alternative for outlier evaluation, enriching the analytical repertoire for researchers and practitioners by foregrounding the interpretative value of outliers within complex, real-world datasets. This paper also is a statement against the dominance of normality in celebration of the luminary and the lunatic alike.
\end{abstract}



\end{frontmatter}


\section{Introduction}
In the design of social robots aimed at mitigating negative moods in individuals with sub-clinical depression, we encountered a notable challenge: the interventions predominantly benefited those with the most severe symptoms, while individuals with less pronounced symptoms, who followed a normal distribution, exhibited little to no responsiveness to robotic assistance \cite{10.1145/335191.335388}. This observation prompted us to question, irrespective of the measurement scale employed, the magnitude an outlier must reach to become receptive to treatment. Furthermore, we pondered the threshold of severity an ailment must attain to warrant the initiation of medication, the consideration of surgical intervention, or to single out the child that needs remedial teaching.

Outlier analysis proves pertinent in numerous other scenarios as well; \cite{10.1145/3381028,SMITI2020100306} highlight its application in detecting cyber-attacks within networks and identifying mechanical malfunctions due to faulty industrial equipment, intrusion detection, fraud prevention, and the diagnosis of medical irregularities. Furthermore, outlier analysis is important in recognising anomalies within wireless sensor networks (the `glitch') and in monitoring deviations in urban traffic patterns (ibid.). On a more positive note, outlier analysis can also assess the level of exceptional performance necessary to qualify an individual as possessing remarkable talent, justifying the awarding of prizes or other accolades.

With \cite{SMITI2020100306}, the outliers we are probing are not incorrect data types, erroneous data values, or missing values. We are not interested in measurement or recording errors, mis-reporting, or sampling error (ibid.). The outlaying values we address may be exceptional but they are meaningful values \cite{SMITI2020100306}.

Various methods for detecting outliers are available, and our approach aligns with the so-called nearest-neighbour-based techniques \cite{10.1145/3381028}, although we do not employ $k$-NN \cite{1053964} in any manner. The approaches most closely related to ours are those of \cite{10.1214/aoms/1177729885} and \cite{10.1214/aoms/1177729747}. Yet, both Grubbs' test and Dixon's $Q$ rely on normally distributed data to estimate the outlier. Grubbs expresses an outlier in terms of means and standard deviations, making his estimator sensitive to deviations from normality. Dixon’s $Q$ does not consider the entire data set; instead, it calculates the difference with the value closest to the outlier, like Grubbs, assuming a symmetric Gaussian distribution. Dixon’s estimator will yield the same results whether the majority of the data is normally distributed with a single distant outlier or forms a uniform histogram while having that one outlier. Dixon’s $Q$ is ineffective if there are clustering effects in the data, as it disregards local density variations.

Our proposed $h^*$ ($h$-star) test statistic for outlier evaluation does not presuppose normally distributed data. Our approach is parametric, frequentist, based on proximity, and can be computed using mean values, although this is not a requirement; $h^*$ can be derived directly from the raw data without necessitating the calculation of means and standard deviations.

As indicated by \cite{10.1145/3381028, 10.1145/1541880.1541882} were the first to discern point outliers from collective outliers. A point outlier refers to a singular data point that significantly diverges from the remainder of the dataset. Collective outliers, on the other hand, consist of a group of data points that seem unusual when compared to the entire dataset, although each point within the group may not individually be considered an outlier. Point outliers can be further categorised into local outliers and global outliers. The identification of local outliers \cite{10.1145/335191.335388} depends on the distinctive differences, like variations in neighbourhood density, between the outlier and its closest neighbours, whereas global outliers are concerned with the disparity in relation to the entire dataset. Also see \cite{SMITI2020100306} on this matter. Our $h^*$ pertains to global outliers.

In conventional statistical analysis, box plots, for example, are frequently employed to visualise data distributions and identify potential outliers. These outliers are typically considered deviations from normality, and the conventional approach often involves their removal to maintain the `integrity' of that desired normality. However, this traditional perspective may overlook the intrinsic value of outliers, which can represent unique phenomena worthy of further investigation (cf.~a Higgs boson or genius-level performance).

In this paper, our focus shifts from the conventional aim of eliminating outliers from a dataset to examining them as individual phenomena of interest in their own right. Our proposed $h^*$ is a test statistic to evaluate the distinctiveness of an outlier much like Student's $t$ is for two means \cite{87464e34-37bc-3288-807c-e421e5a0d7a6}. Rather than assessing normal distribution characteristics, as, for instance, box plots, interquartile range (IQR), or (modified) $z$-scores do, the test statistic $h^*$ comes with a measure of confidence regarding the extremity of a value compared to its group counterparts, where normality is not necessarily required. This approach allows us to ascertain whether an outlier is stably exceptional or part of the normal data distribution with some values missing at the tail (cf.~Dixon’s `gap'). In addition, the generalised $h^*$ test statistics incorporate contextual individual or pair-wise features of the members and the environment specifications and variable exponent, which distinguishes itself from the traditional Grubbs test, allowing for enriched domain descriptions, optimisation of test power and modelling of subjective inference.

The $h^*$ test statistic underscores the nuanced analytical strategy, wherein data are analysed both with and without outliers to discern the differences. Sometimes, the results with outliers included do not change the pattern of results found without the outliers. Removing suspected outliers up front would be unwarranted in spite of disrupting normality. We will show that the $h^*$ test statistic reliably recognises that some values identified as outliers by conventional techniques may not be statistically significant outliers, whereas others may indeed represent stable extraordinary deviations. With $h^*$, we can test the validity of the null hypothesis; $H_0$ being that there is not strong enough evidence to assume the outlier. By employing $h^*$, we can assert that certain extreme values, traditionally flagged as outliers, do not warrant exclusion, thereby enriching our understanding of the underlying phenomena that produced them. This approach not only challenges the conventional methodology but also enhances our ability to appreciate the complexity and diversity inherent in the world that is reflected in our datasets, a world that is not necessarily `normal' in a mathematical sense.

The $h^*$-test statistic is not for the detection of outliers like box plots, for example. It is not a descriptive tool but a means of assessing the outlier as an object of investigation in and of itself. In the remainder of this paper, we will first outline $h^*$ in its simplest and raw form to show its inner workings. Then we will reformulate $h^*$ mathematically and provide confidence intervals, the way to calculate degrees of freedom, and a table of significance. We will demonstrate the working of $h^*$, using a real data sample, containing extreme values.

\section{The $h^*$ test statistic for outlier evaluation} \label{sec:h-formulation}
The $h^*$ test statistic for outlier evaluation expects data measured on interval to ratio level. Rating scales from 6 points up have still some ordinality in them but may still suffice for reliability as well as convergent and divergent validity \cite{chomeya2010quality, PRESTON20001}. The $t$ statistic

\begin{equation} \label{eq:1}
    \text{If } t_{\text{(df)}} = \frac{\overline{X} - (\mu = x^*)}{\sigma / \sqrt{n}} \rightarrow p < \alpha, \text{ then:}
\end{equation}
is used for outlier detection, not assessment. To determine whether an outlier is different enough from average behaviour ($\bar{x}$), (\ref{eq:1}) is sufficient but not necessary and could be replaced by other techniques such as box plots. Note that in (\ref{eq:1}), the one-sample $t$-test assumes no hypothetical $\mu$, but takes on the outlier value $x^*$ as $\mu$.

If the outlier is detected through (\ref{eq:1}), we may employ
\begin{definition} \label{eq:3}
For $n$ i.i.d. random variables $\mathbf{X}=(X_1, \dots, X_n)^\intercal$, define the statistic

\begin{equation*} 
h^* = \sqrt{\frac{\left.\displaystyle\sum_{k=1}^{n}(X_k - X^*)^2 \middle/ (n-1)\right.}{\left.\displaystyle\sum_{\substack{i>j \\ X_i, X_j \neq X^*}}(X_i - X_j)^2 \middle/ \displaystyle\frac{(n-1)(n-2)}{2}\right.}} \\
\end{equation*}
where $X^* = \max\{\mathbf{X}\}$, the random variable as the candidate outlier. 
\end{definition}
\begin{lemma} \label{lem:h-formula-alt}
$h^*$ can be expressed as 
\begin{equation*} 
h^* = \sqrt{\frac{n-2}{2}} \sqrt{\frac{\displaystyle{\sum_{k=1}^{n}X_k^2 - 2X^*\sum_{k=1}^{n}X_k + n{X^*}^2}}{\displaystyle{(n-1)\left(\sum_{k=1}^{n}X_k^2 - {X^*}^2\right) - \left(\sum_{k=1}^{n}X_k - X^*\right)^2}}}.
\end{equation*}
\end{lemma}
The proof of Lemma~\ref{lem:h-formula-alt} is given in Appendix~\ref{sec:h-eq-proof}.

\begin{lemma} \label{lem:hRange}
The range of $h^*$ is $h^* \in \left[\frac{1}{\sqrt{2}}, \infty\right)$.
\end{lemma}
The proof of Lemma~\ref{lem:hRange} is given in Appendix~\ref{sec:h-full-proof}.

\begin{collorary} \label{lem:reverse}
In the case of $X^* = \min\{\mathbf{X}\}$, Definition~\ref{eq:3} can be applied with the transformation $\mathbf{Y} = -\mathbf{X}$ to reverse the order.
\end{collorary}
The $h^*$-test is a statistical, parametric test for global outliers in one-dimensional homogeneous data. As depicted in (\ref{eq:1}), we assume an outlier is either an unusually small or large data point. Therefore, the $h^*$-test is a one-tail test. Definition~\ref{eq:3} shows that $h^*$ is the ratio of the root-mean-square (rms) pairwise distances between the member of ordinary data and the candidate outlier (signal) to the rms pairwise distances among all members of ordinary data (i.e.,~excluding the candidate outlier), which are irrelevant to outlier inference (noise). In other words, the expression is the ratio of the measure of dispersion from $X^*$, an analogy of standard deviation, to the measure of pairwise difference of the ordinary data, or equivalently, the characteristic distance from the outlier normalised by the characteristic ordinary data pairwise difference. The numerator contains $n-1$ differences (the term $X_k=X^*$ vanishes). The denominator takes all pairwise combinations without $X^*$, thus containing $(n-1)(n-2)/2$ terms. Generally, the higher the value of $h^*$ for a sample of a given size $n$, the greater the distance of the candidate outlier from the ordinary data.

\begin{definition} \label{def:h}
The associated test statistic $\tilde{h} \in \left[\frac{1}{\sqrt{n-2}}, \infty\right)$ is 
\begin{equation*}
    \tilde{h}^* = \sqrt{\frac{2}{n-2}}h^*.
\end{equation*}
\end{definition}
\begin{figure}[h!]
    \centering
    \includegraphics[width=0.5\linewidth]{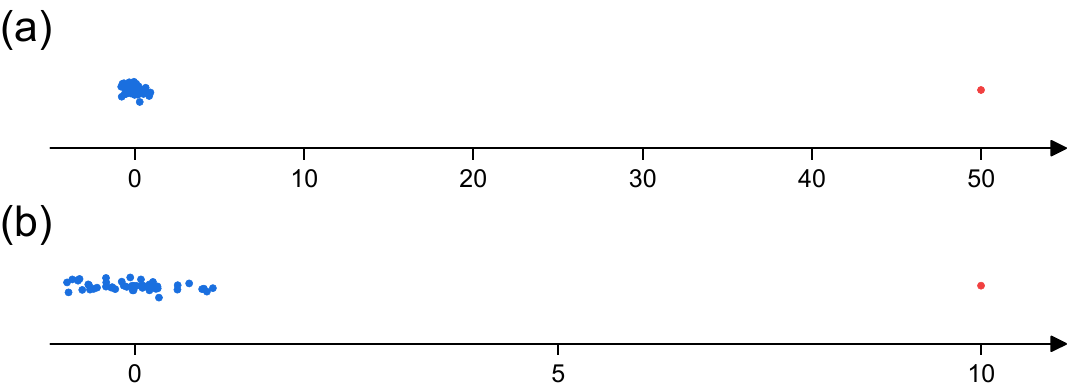}
    \caption{An illustration of $h^*$ with the candidate outlier (red) having different distances with the ordinary data (blue). The candidate outlier in (a) is far from the ordinary data point, making the ordinary data localised with small pairwise distances compared to the distance to the candidate outlier. In contrast, the outlier candidate in (b) has a smaller distance from the same ordinary data. The `zooming' results in an apparent wider spread of the ordinary data and more varied distances.}
    \label{fig1}
\end{figure}

Despite the difference from $h^*$ with the scale factor $\sqrt{\frac{n-2}{2}}$, $\tilde{h}^*$ accounts for the effective dominance of dimensions in the difference space (i.e.,~difference between the ordinary data and the candidate outlier, see (\ref{eq:17})) as an alternative perspective. As illustrated in Figure~\ref{fig1}, $\tilde{h}$ is small (down to $\frac{1}{\sqrt{n-2}}$) if the dataset sees relatively variable distances from the candidate outlier in the length scale between the ordinary data and the candidate outlier, and $\tilde{h} \to \infty$ if the dataset sees a negligible distance variation from the candidate outlier (like any point away from the origin out of the $\mathrm{\delta}$-distribution). Notably,

\begin{lemma} \label{lem:h-invariance}
$h^*$ and $\tilde{h}^*$ are invariant under linear transformations.
\end{lemma}
The proof of Lemma~\ref{lem:h-invariance} is depicted in Appendix~\ref{sec:h-invariance}.

In an example with four values of which one is an outlier, suppose the values 3, 4, and 5 represent the distribution and 8 the outlier value. Then Definition~\ref{eq:3} simply states:
\vspace{1em}

Numerator (signal) =
\begin{alignat*}{3}
    & n_{\text{pair1}}: \quad & (3-8)^2 & = 25 \\
    & n_{\text{pair2}}: \quad & (4-8)^2 &= 16 \\
    & n_{\text{pair3}}: \quad & \underline{+ \quad (5-8)^2} & \underline{\;= \ \:9\phantom{()}} \\
    & {\text{rms of outlier difference}}\quad && = \sqrt{50/3} = 4.08
\end{alignat*}
\vspace{1em}

Denominator (noise) =
\begin{alignat*}{3}
    & N_{\text{pair1}}: \quad & (3-4)^2 &= 1 \\
    & N_{\text{pair2}}: \quad & (3-5)^2 &= 2 \\
    & N_{\text{pair3}}: \quad & \underline{+ \quad (4-5)^2} & \underline{\;= 1\phantom{()}} \\
    & {\text{rms of inlier difference}}\quad && = \sqrt{4/3} = 1.15
\end{alignat*}
\vspace{1em}
Thus, $h^* = 4.08/1.13 = 3.54$. 

\begin{figure}[h!] 
    \centering
    \begin{minipage}[t]{0.48\textwidth}
        \centering
        \includegraphics[width=\linewidth]{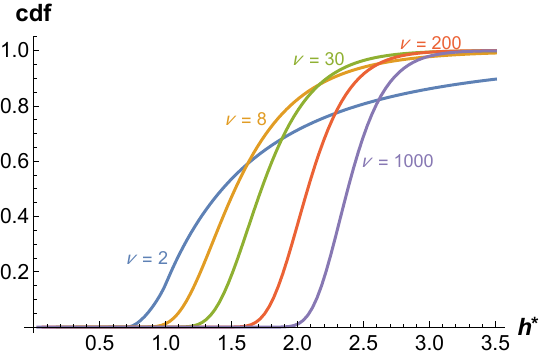}
        \vspace{0.5em}
        \small
        (a) Normal distribution: $X_k \sim N(\mu, \sigma^2)$
    \end{minipage}%
    \hfill
    \begin{minipage}[t]{0.48\textwidth}
        \centering
        \includegraphics[width=\linewidth]{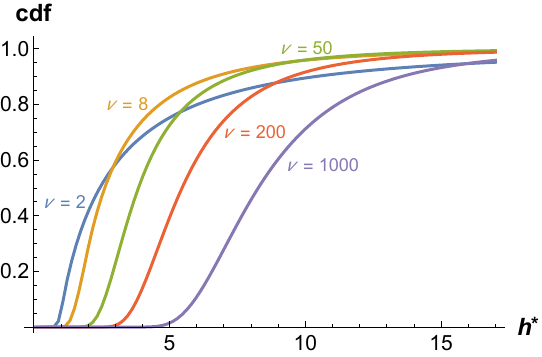}
        \vspace{0.5em}
        \small
        (b) Log-normal distribution: $\ln X_k \sim N(0, 1^2)$
    \end{minipage}
    \vspace{1em}
    \caption{The cumulative distribution functions (cdf) of $h^*$ of different degrees of freedom $\nu$ based on the (a) normal distribution and (b) log-normal distributions of standardised random variables.}
    \label{fig2}
\end{figure}

Obviously, the $h^*$ value does not mean much if not held against a confidence level and a measure of significance. To determine confidence and significance level, the probability density function (pdf) of $\tilde{h}^*$ is introduced:

\begin{theorem} \label{thm:h-pdf}
Assume an unbounded support of $X_k$, i.e.~$X_k \in (-\infty, \infty)$. Let $f_X(x)$ be the probability density function (pdf) of $X_k$ and $\mathbbm{1}\!(q)$ be the indicator function which gives 1 if the logical statement $q$ is true, and 0 if false. The pdf of $\tilde{h}^*$ is
\begin{equation} \label{eq:4}
    \tilde{f}\!\left(\tilde{h}^*\right) = \int_0^{\frac{\pi}{2}} \cdots \int_0^{\frac{\pi}{4}} \int_0^\infty \int_{-\infty}^\infty f_{\mathrm{hs}}\!\left(u_1, R, \tilde{h}^*, \varTheta_2, \dots, \varTheta_{n-2}\right) \,\mathrm{d}u_1 \,\mathrm{d}R \,\mathrm{d}\varTheta_{n-2} \dots \mathrm{d}\varTheta_2,
\end{equation}
where
\begin{equation} \label{eq:5}
\begin{aligned}
& \quad f_{\mathrm{hs}}\!\left(u_1, R, \tilde{h}^*, \varTheta_2, ..., \varTheta_{n-2}\right) \\
&= n! f_X\!\left(u_1\right) f_X\!\left(u_1 - \frac{R}{\tilde{h}^*}\sqrt{\frac{(n-1)\tilde{h}^{*2}-1}{n-1}}\right) \\
& \times \prod_{k=3}^{n-1} f_X\!\left(u_1 - \frac{R}{\sqrt{n-1}\tilde{h}^*} \left[\prod_{j=2}^{k-2} \sin\varTheta_j\right] \cos\varTheta_{k-1}\right)  \\
& \times f_X\!\left(u_1 - \frac{R}{\sqrt{n-1}\tilde{h}^*} \prod_{j=2}^{n-2} \sin\varTheta_j \right) \\
& \times \mathbbm{1}\!\Biggl(\left( \tilde{h}^* \ge \frac{1}{\sqrt{n-2}} \right) \wedge \left(\sqrt{(n-1)\tilde{h}^{*2}-1} \ge \cos\varTheta_2 \right) \wedge \left(\bigwedge_{k=2}^{n-3} \cot\varTheta_k \ge \cos\varTheta_{k+1}\right)  \\
& \qquad \wedge (\cot\varTheta_{n-2} \ge 1) \Biggr) \\
& \times \frac{R^{n-2}}{\left(\sqrt{n-1}\tilde{h}^*\right)^{n-3}} \prod_{k=2}^{n-3} \sin^{n-k-2}\varTheta_k \cdot \frac{1}{\tilde{h}^*\sqrt{(n-1)\tilde{h}^{*2}-1}}.
\end{aligned}
\end{equation}
\end{theorem}
 The proof of Theorem~\ref{thm:h-pdf} is given in  Appendix~\ref{sec:h-full-proof}. According to (\ref{eq:21}), the mean represented by the sum subtracts one degree of freedom from the $n-1$ variables given the constraint that the sum of differences from their mean is zero. Therefore, the test statistic has $\nu = n-2$ degrees of freedom. In practice, Monte Carlo simulations are used to generate the $h^*$ distributions based on the distribution of $X_k$ instead of evaluating the $n$-dimensional integral whose complexity increases with dimensionality. As an example, the $h^*$ distributions for $X_k$ being normally and log-normally distributed are depicted in Figure~\ref{fig2}, and the single-tail confidence table in Table~\ref{tab:normal-table} and \ref{tab:log-normal-table}. For every degree of freedom (sample size), $10^8$ simulations were performed, where the values of $h^*$ were calculated from the generated random variates and binned in intervals of $0.0025$. In principle, the $h^*$-test can be performed on various distributions by simulating the corresponding $h^*$ distributions. 

\section{Demonstration of the $h^*$ test statistic with empirical data} \label{sec:h-test-demo}
The data we used to evaluate our $h^*$ test statistic were published in the Supplementary Materials of \cite{robotics10030098}. These data were sampled from voluntary participants ($N=45$; $M_{\text{age}} = 24.9$, $SD_{\text{age}} = 3.29$; 55.6\% female, 44.4\% male; Chinese nationality), who were randomly assigned to a between-subject experiment of self-disclosure in a Robot ($n=24$; 54.2\% female) versus Writing condition ($n=21$; 57.1\% female) after negative-mood induction. These participants scored Likert-type items on a structured questionnaire, using rating scales [1--6], ranging from totally disagree to totally agree, respectively. These are semi-interval scale values.

The dataset may undergo a trial of outlier detection. The null hypothesis of the $h^*$-test is that the greatest (or smallest by focusing the greatest of the negated data) $n'$ data point(s) are not \emph{collectively} the outliers for the dataset of size $n$ that follows the prior distribution $D$. The tail characteristics of the prior distribution is particularly relevant to the sample extrema, impacting the inference of whether the selected data points are outliers. Assuming that no outlier has a value exclusively bounded by the range of the ordinary values, the $n'$ extrema are separated from the ordinary values. The prior distribution $D$ for the ordinary dataset of size $n-n'$ is justified by an appropriate goodness-of-fit hypothesis test. Then, the candidate outliers are incorporated into the ordinary dataset subsequently, each forming a dataset of size $n-n'+1$ to test against $h^*$. From a frequentist's point of view, the null hypothesis is then rewritten as the disjunction of every data point of the $n$ maxima (or minima) being not an outlier for the dataset with a prior distribution $D$, i.e.,~an intersection-union test (IUT). The extrema are regarded as outliers only if all outlier candidates lead to rejection of their respective null hypotheses. The above procedures are iterated for different candidate outlier combinations for correct identification.

If $D$ is found not to fit the ordinary data, it may be that the prior knowledge about the dataset is invalid, or the ordinary data misses the tail characteristics too significantly to represent the distribution correctly. Re-examination of the prior distribution or dataset may be advised.
\pagebreak

\begin{figure}[h!] 
    \centering
    \makebox[\textwidth][l]{(a)}
    \includegraphics[width=\textwidth]{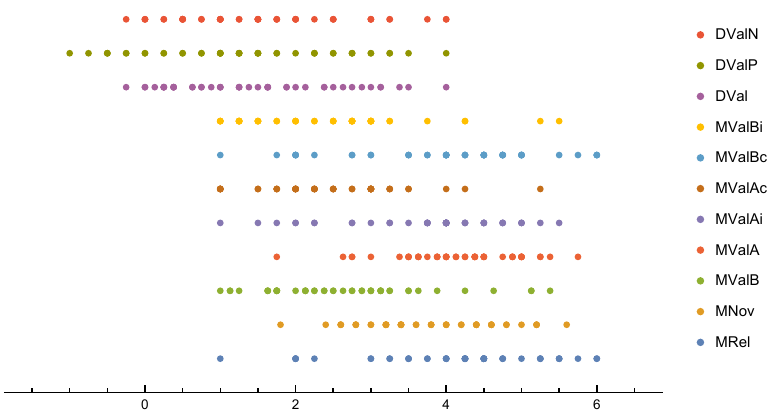}
    \vspace{1em}
    \makebox[\textwidth][l]{(b)}
    \includegraphics[width=\textwidth]{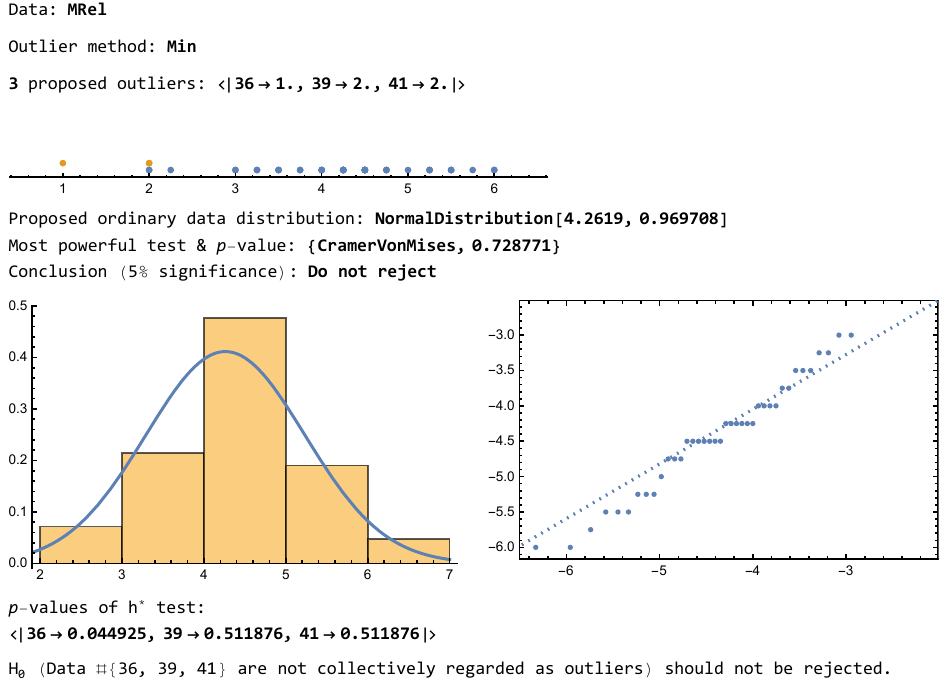}
    \caption{
        (a)~One-dimensional scatter plot of mean (prefix 'M') and difference (prefix 'D') scores of various dataset variables. The legenda to the variables are depicted as follows: ValB--—Valence before robot interaction or writing, ValA—Valence after robot interaction or writing, Rel—--Relevance, Nov—Novelty (covariate), B--—before treatment, A—--after treatment, c—--indicative item, i--—counter-indicative item, p--—positive, n--—negative. (b)~Example output of the $h^*$-test.
    }
    \label{fig3}
\end{figure}

The mean and difference scores of the empirical data were extracted to examine potential outliers (Figure~\ref{fig3}(a)). For each dataset item, both ends were tested with trials of up to three candidate outliers, carrying out $11 \times 2 \times 3 = 66$ trials. The analysis identifies the candidate outlier(s), fits the ordinary data with an empirical normal distribution, followed by a distribution test and a Q-Q plot, and evaluates the $p$-value of $h^*$ based on the size of the sample (Figure~\ref{fig3}(b)). The results are tabulated in Table~\ref{tab1}. Three sets of outliers were found under the 5\,\% significance level: the minimum of  MValA, the maximum of MValAc, and the maximum two of MValBi. The extrema of MValA and MValAc are notably distant from the rest of the data. For MValBi, the two greatest data points form a cluster distant from the ordinary data. Treating only the maximum as the candidate outlier leads to an unreasonably small distance from the next maximum, decreasing $h^*$ and failing at rejecting the null hypothesis, but only if treating them as outliers collectively. Whereas the left tail of MRel contains sparse data, their inter-data distances relative to the ordinary data are only moderate, resulting in a greater $p$-value when treating any of the consecutive three as outliers.

\section{Power of $h^*$-test}
\begin{figure}[h!]
    \centering
    \includegraphics[width=0.9\linewidth]{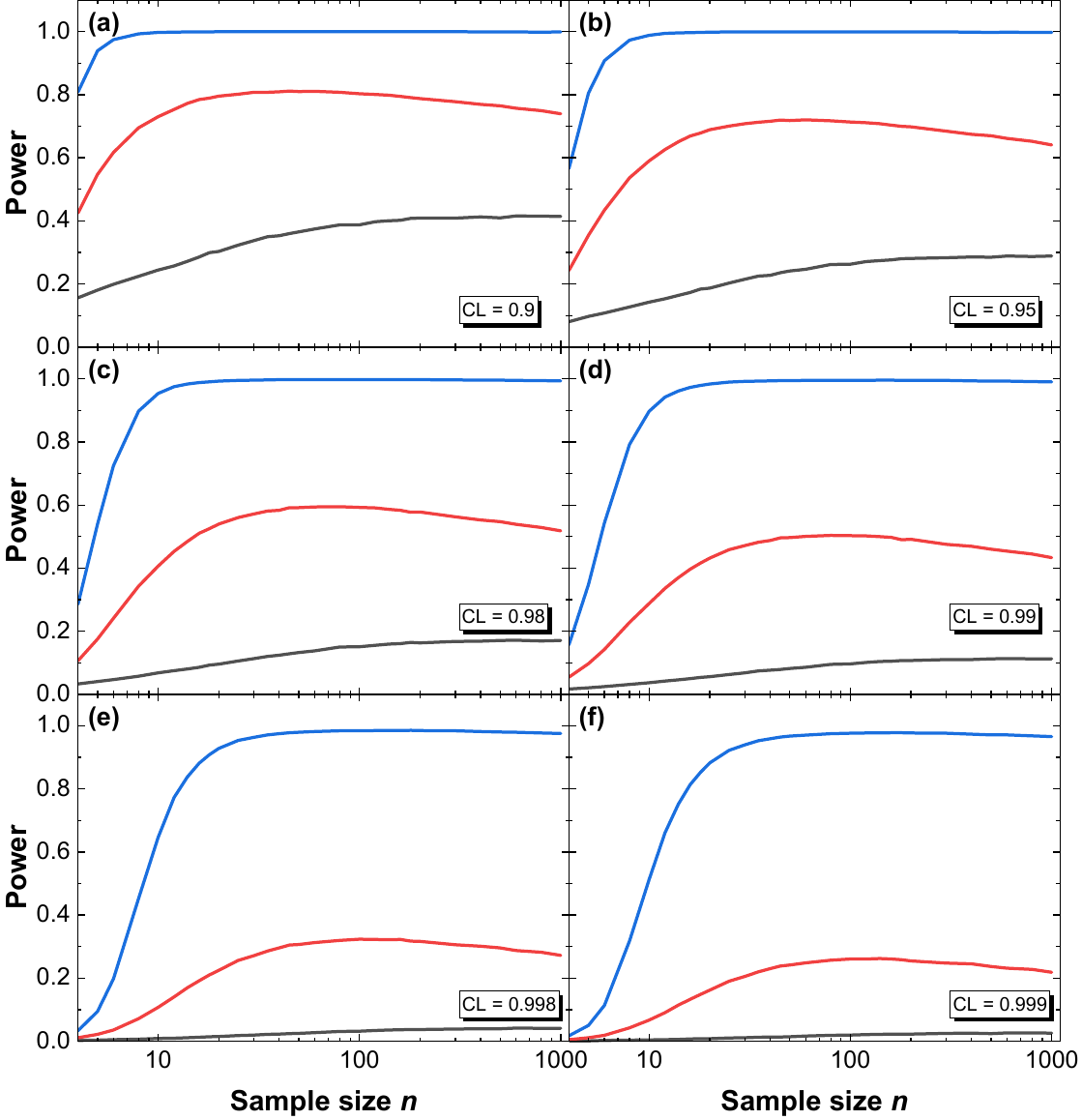}
    \caption{Power of $h^*$-test for i.i.d. normally distributed samples with one outlier of mean shift of size effects of 1.7 (black), 3.7 (red) and 6.6 (blue) under different confidence levels (CLs).}
    \label{fig4}
\end{figure}

Figure~\ref{fig4} displays the simulation results of the power of the $h^*$-test of three effect sizes, indicating the mild (1.7), moderate (3.7) and strong (6.6) outliers. The simulation was constructed from i.i.d. standardised normally distributed samples with one outlier modelled by a mean shift characterised by the effect size. The value of power is the ratio of the number of simulations with $h^*$ in the critical region to the total number of simulations ($10^5$). Generally, the power increases with the effect size and sample size. The boost with sample size increases with the effect size. The $h^*$-test is confidently accurate with $\sim 10$ samples for the large effect size. The power reaches above 60\% for the moderate outlier and up to 30\% for the mild outlier under a 95\% confidence level. The power has a gentle falloff at large sample sizes (except for the small effect sizes, where the displayed sample size is not sufficiently large for such observation). This can be attributed to the occurrence of more extreme events in the ordinary data.

\section{Effect of the choice of candidate outlier on the $h^*$-test} \label{sec:choice}
The IUT nature accounts for the fact that the choice of the candidate outliers affects the inference. The expression of $h^*$ and $\tilde{h}$ can be transformed into one in terms of the differences from the maximum or candidate outlier ($U_k = X_{(n)} - X_{(k-1)}$ where $\{X_{(k)}\}$ are $\{X_k\}$ in ascending order, $k>1$), coined as the difference space, involving the ratio between the square of sums ($Q^2$) and the sum of squares ($R^2$) (\ref{eq:21}). Noting the invariance of $h^*$ under linear transformations, the reciprocal of this ratio is analogous to the inverse participation ratio (IPR) \cite{Johnston_1990} by normalising $U_k$ with the L1 norm:

\begin{equation} \label{eq:6}
    U'_k = \frac{U_k}{\displaystyle\sum_{j=2}^{n} U_j}
\end{equation}
so that

\begin{equation} \label{eq:7}
    \frac{\displaystyle\left(\sum_{k=2}^{n} U'_{k}\right)^{2}}{\displaystyle\sum_{k=2}^{n} {U'_{k}}^{2}} = \frac{1}{\displaystyle\frac{\sum_{k=2}^{n} U_{k}^{2}}{\left(\sum_{j=2}^{n} U_{j}\right)^{2}}} = \frac{\displaystyle\left(\sum_{k=2}^{n} U_{k}\right)^{2}}{\displaystyle\sum_{k=2}^{n} U_{k}^{2}} = \frac{Q^2}{R^2}.
\end{equation}

Despite that IPR is a concept in quantum mechanics, the analogy results solely from the formulation similarity: for a quantum system of discrete states with non-negative probability amplitudes $\psi_i$ ($i=1, \dots, N$) such that $\sum_{i=1}^{N} |\psi_i|^2 = 1$, IPR $= \sum_{i=1}^{N} |\psi_i|^4$ is a measure of the localisation of the quantum state. A completely localised state has IPR $= 1$, and a uniformly spread state has IPR $= 1/N$. Here, $Q^2/R^2$ can be interpreted as the reciprocal of IPR, i.e.,~the effective number of participating differences, by associating $|\psi_k|^2$ with $U_k$. A participating difference is a sufficiently significant difference to signify the selected maximum does not probably belong to the ordinary population. Intuitively, $U_k$ can be classified into participating and non-participating. Ordinary data far from the selected maximum are regarded as participating. Therefore, the expression of $\tilde{h}$ in (\ref{eq:21}) may be interpreted as the square root of the so-called inverse non-participating ratio (INR) of $U_k$. The less the effective number of non-participating differences (1/INR), the more restricted the data can ‘explain away’ the extreme point without calling it an outlier. The resulting higher values of $h^*$ and $\tilde{h}$ indicate a stronger inference to reject the null hypothesis. Therefore, $h^*$ and $\tilde{h}$ are regarded as measures of \emph{negated signal detection}, i.e.~the explicit construction does not detect how likely a data point is an outlier but how likely to reject a point not to belong to the ordinary data, as typical null hypotheses of frequentist outlier detection are formalised.

For a candidate outlier far from the spread of the ordinary data with respect to the scale of the ordinary data pairwise distances, the variance of the normalised $U_k$ is generally small, indicating comparable $U_k$ participation and thus small IPR. Specifically, $\tilde{h} \to \infty$ as $1/\text{IPR} = Q^2/R^2 \to n-1$, signifying rejection of the null hypothesis. Conversely, the variance of $U_k$ increases with decreasing normalised distance between the candidate outlier and the ordinary data, resulting in small $U_k$ for ordinary data points neighbouring the candidate outlier, and large $U_k$ for those at the opposite end. This increases the IPR and decreases the test statistics. In particular, $\tilde{h}^* \to 1/\sqrt{n-2}$ as $Q^2/R^2 \to 1$ (a particular $U_k$ dominates). This formulation provides a fuzzy way to quantify the membership of the selected maximum in the ordinary data (or not) by considering the \emph{relative positions of all} data points. In other words, the choice of candidate outliers determines the distance between the outliers and the ordinary data. If there are two points at proximity so far away from the ordinary data, but only one of them is identified as the outlier, it may be too close to the ordinary data to be justified as an outlier, unless both points are regarded as outliers. Therefore, the $h^*$-test is a proximity-based approach that considers both the neighbourhood and clustering. Instead of $k$-NN or neighbourhood within a pre-specified radius (up to the iteration range), $h^*$-test looks into the overall topology of the data distribution and does the inference by determining the effective number of differences from the candidate outlier that are considered significantly far or near to it. The approach is distinct from the traditional Dixon's $Q$ test (nearest-neighbour-based) or Grubb's test (test for single outliers with presumed radius threshold).

\section{Effect of sample size on the values of $h^*$}
\begin{figure}[H]
    \centering
    \begin{minipage}{0.48\textwidth}
        \includegraphics[width=\linewidth]{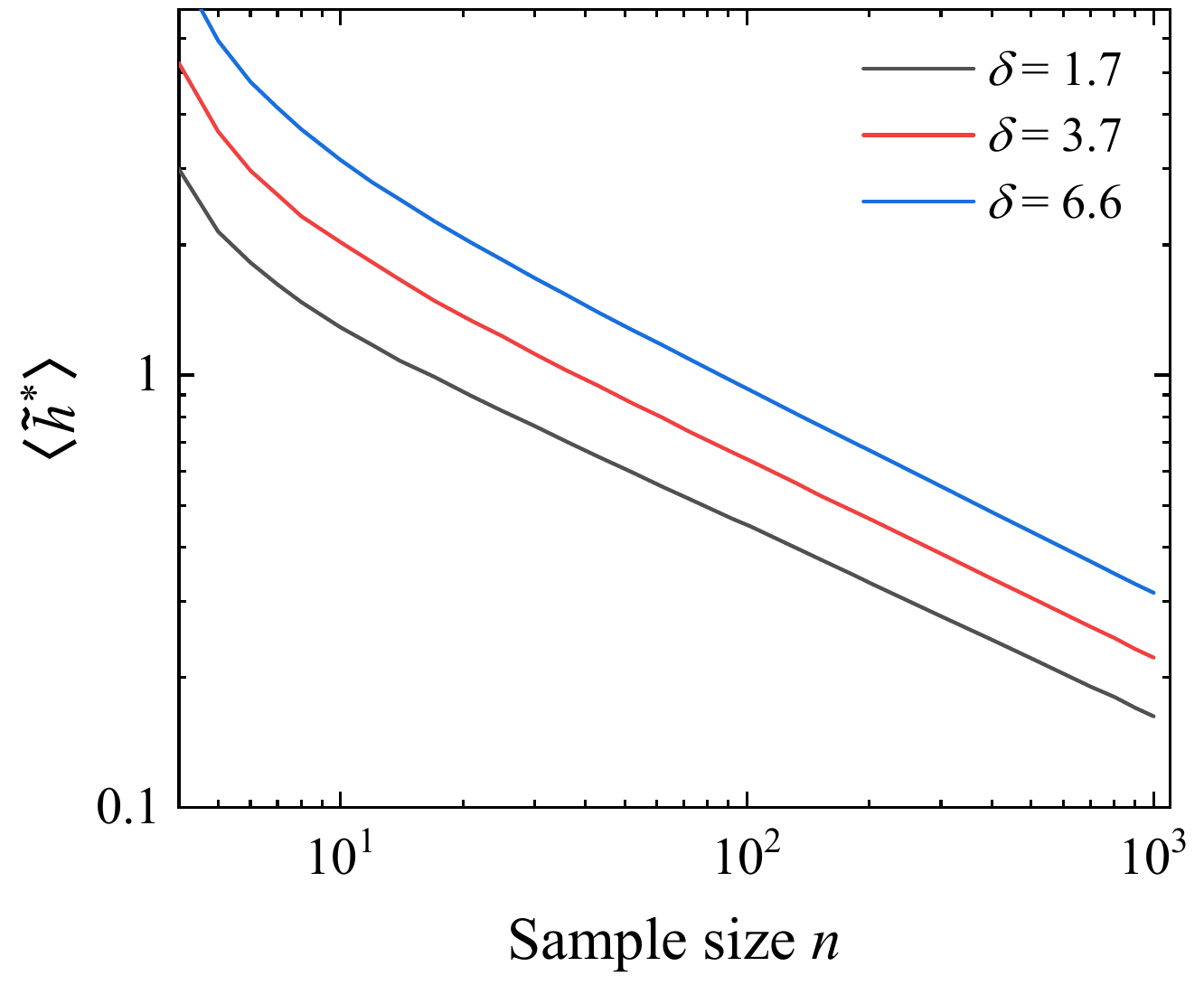}
        \centering
    \end{minipage}
    \hfill
    \begin{minipage}{0.48\textwidth}
        \includegraphics[width=\linewidth]{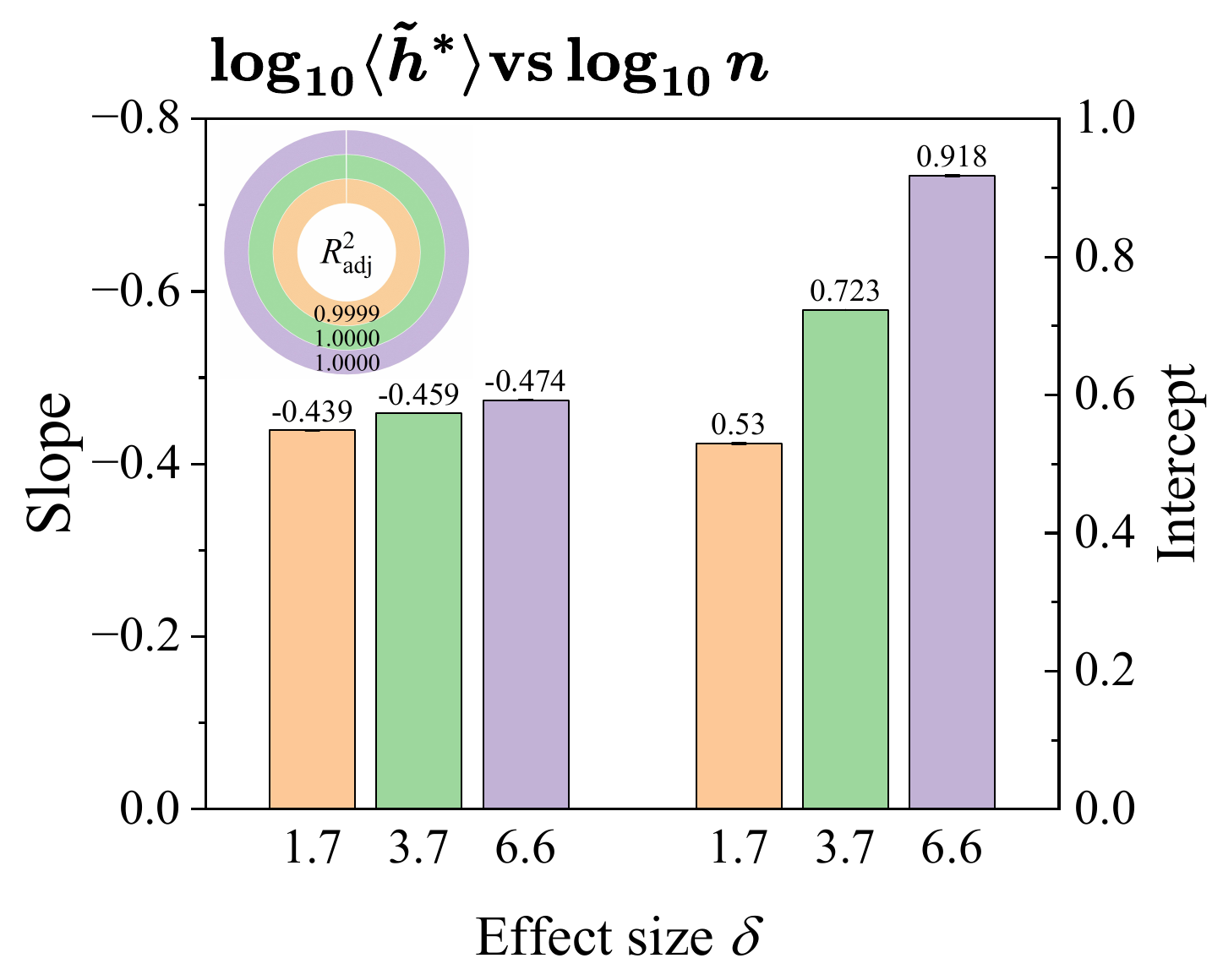}
        \centering
    \end{minipage}
    \vspace{0.5cm}
    \begin{minipage}{0.48\textwidth}
        \includegraphics[width=\linewidth]{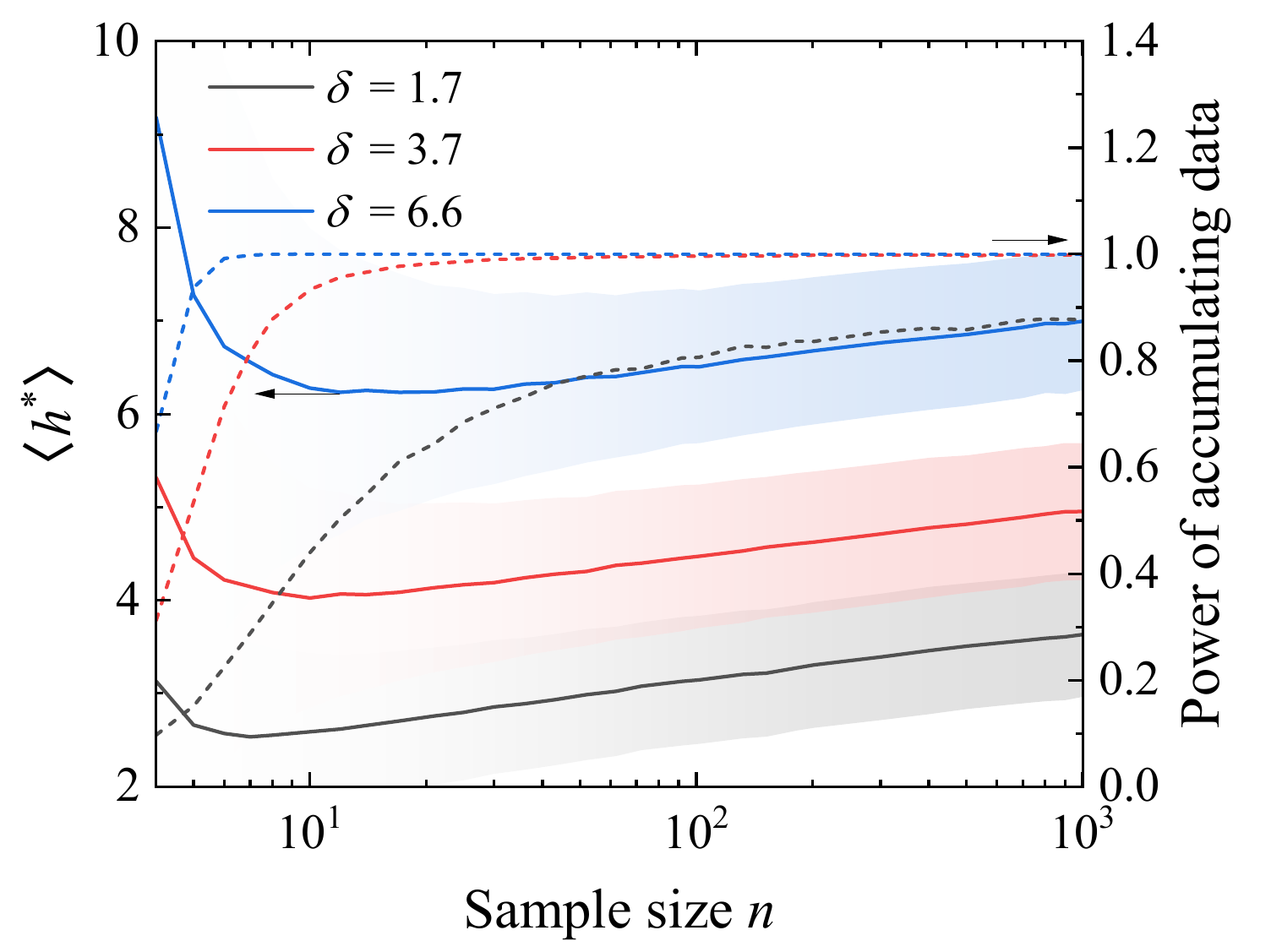}
        \centering
    \end{minipage}
    \hfill
    \begin{minipage}{0.48\textwidth}
        \includegraphics[width=\linewidth]{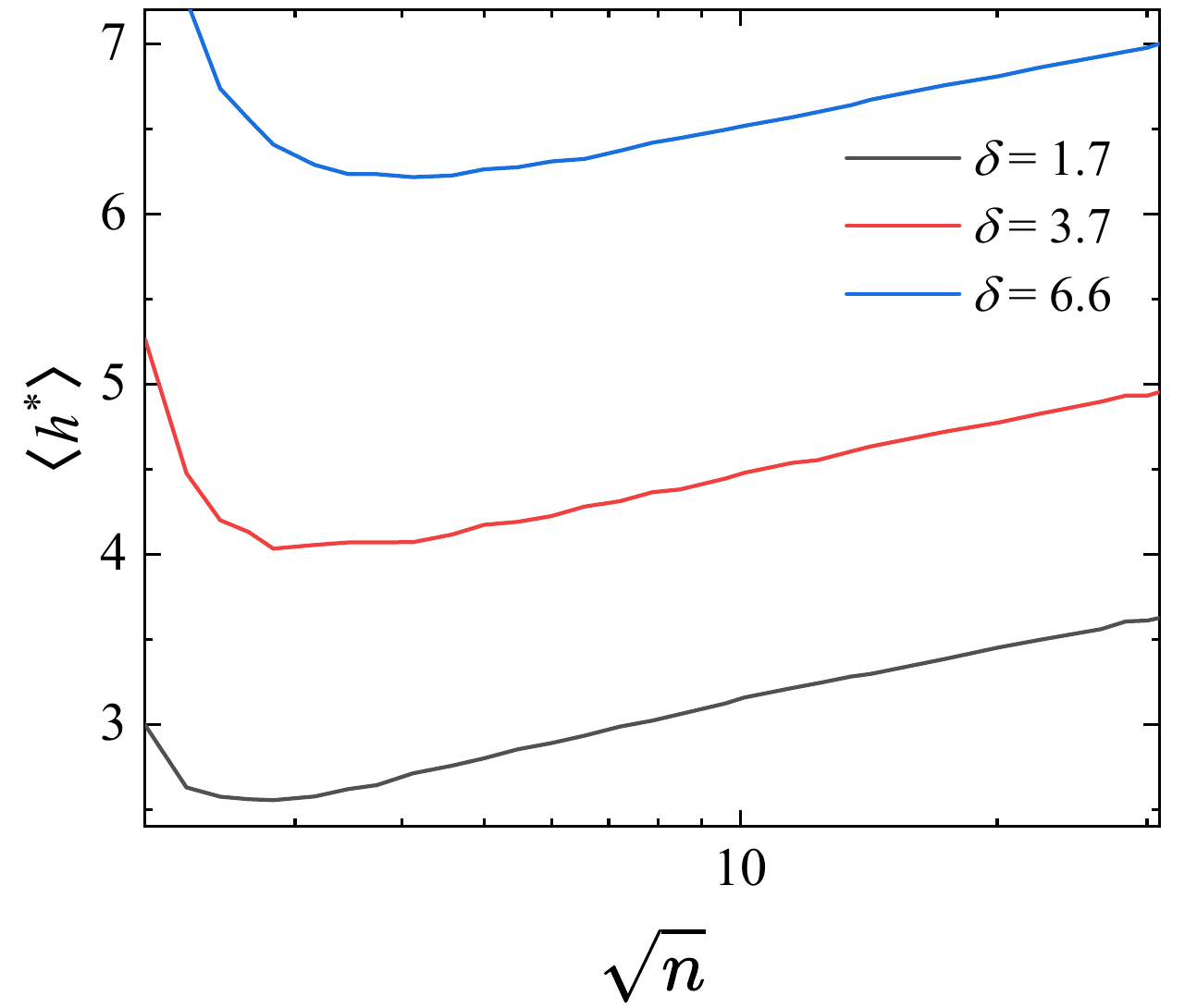}
        \centering
    \end{minipage}
    \vspace{0.5cm}
    \begin{minipage}{0.98\textwidth}
        \includegraphics[width=\linewidth]{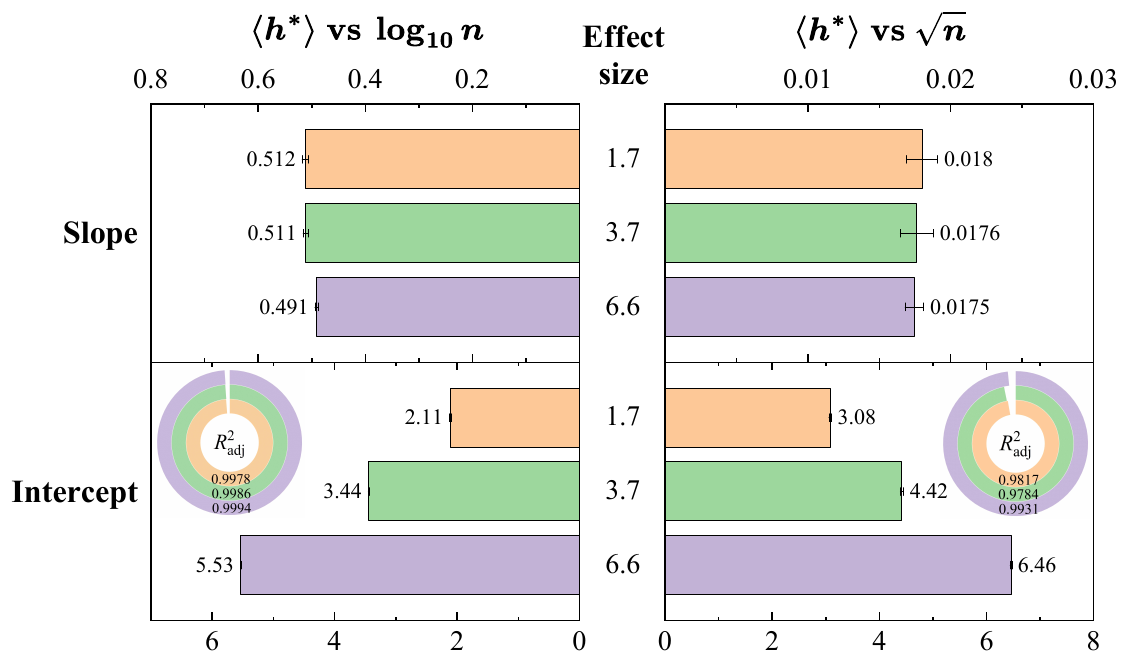}
        \centering
    \end{minipage}
    \caption{Simulation results of the cumulative sample size effect on $h^*$ for outliers of various effect sizes $\delta$. (a) The log-log plot of the ensemble average over all trials of the test statistics $\langle h^* \rangle$ vs sample size (including the outlier) $n$ and (b) the summary of linear regression for (a) in the linear region, showing the slope (left) and y-intercept (right) together with the adjusted $R^2$ (inset) for each effect size. (c) The linear-logarithmic plots of $\langle h^* \rangle$ (solid lines) with standard deviations (filled) and the power of inference based on accumulating observations under the 95 \% confidence level (dashed lines). (d) The plot of $\langle h^* \rangle$ vs $\sqrt{n}$. (e) Summary of linear regression for (c) and (d) in the linear region, showing the slope (top) and y-intercept (bottom) together with the adjusted $R^2$ (inset).}
    \label{fig5}
\end{figure}

Figure~\ref{fig5} shows the results of the behaviour of $h^*$ from the simulation under the accumulation of normally distributed observations with an outlier, mimicking repeated temporal short-tailed observations with the persistent presence of a `surprise'. In each of the $10^4$ trials, the sample size is modulated by appending sampling points to the current sample to simulate the temporal observation accumulation. The outlier, supposedly invariant with the ordinary distribution, is modelled to be a fixed number with respect to the sample size, originating from the truncated normal distribution with the same variance and mean shift (effect size) $\delta$ so that it is greater than the entire ordinary data of all sample sizes.

The value of $h^*$ decreases at small sample size and then increases proportionally in the linear-logarithmic scale for large samples, with the size effect providing a shift in $h^*$ (Figure~\ref{fig5}(c)). The initial decrease results from the dominating decrease of $\tilde{h}^*$ (Figure~5(a)), i.e.,~the INR, where a more probable and significant addition of extreme observations in the small sample that does not fully represent the data variability signifies the contribution of slight differences from the outlier and the effective number of non-participating differences. However, $\tilde{h}^*$ decreases with the sample size exponentially in the form of $\alpha n^{\beta}$ (Figure~\ref{fig5}(b)) for large samples, with a slowly varying exponent from $-0.44$ to $-0.47$ for the three effect sizes, indicating a diminishing marginal decrease of the INR. This happens because the differences due to rarer occurrences that are close to the outlier are diluted by the mainstream data points that are farther away from the outlier, leading to increasing but gradually saturated participating differences. This results in an increase in $h^*$ in larger samples with the competing sample size effect of $\sqrt{\frac{n-2}{2}}$. Surprisingly, linear regression analysis of Figure~\ref{fig5}(c) reveals that $h^*$ increases with $\log\sqrt{n}$ for all the three effect sizes, where the square root originates from the slope of regression (0.491--0.512 in Figure~\ref{fig5}(c)). In addition, the rate of increase in $h^*$ with respect to $\sqrt{n}$ (Figure~\ref{fig5}(d)) was also found to have a subtle difference of $\approx 3$ \% for the three effect sizes ($\partial \langle h^*\rangle/\partial\sqrt{n} \approx 0.018$) in the range of simulated sample sizes (Figure~\ref{fig5}(c)). The two linear regression models can be related using the Taylor expansion around some arbitrary point $\sqrt{n}=a$ up to the first order: $\log_{10} \sqrt{n} \approx \left(\log_{10} a - \frac{1}{\ln 10}\right) + \frac{1}{a \ln 10} \sqrt{n}$. The constant slope of regression between $h^*$ and $\sqrt{n}$ indicates that the typical range of the shape feature $h^* \sim \sqrt{n}$ and thus the entire region of $\log \tilde{h}$ is invariant with the effect size for the simulated range of effect sizes up to a vertical shift.

As a remark, Figure~\ref{fig5}(c) also shows the power of accumulating observations. Compared to the results in Section~\ref{sec:choice}, there is more power increase with observation accumulation. The power reaches 0.8 at $N \approx 100$ even for mild outliers of the effect size of 1.7, and saturates at $N=30$ and $N=6$ for the effect size of 3.7 (moderate outliers) and 6.6 (strong outliers). This further demonstrates the validity of the $h^*$-test.
\pagebreak

\section{The $h^*$-test in the Bayesian framework} \label{sec:bayesian}
The $h^*$-test can also be applied to the Bayesian framework. Because $\tilde{h}$ and $h^*$ are invariant under linear transformations (see (\ref{eq:32})), the protocol is delivered in a relative scale of the random variables and carried out using the standardised ones. Similarly, presuming that the candidate outliers are the extrema without interactions with each other, they are first separated from the ordinary data. Each set of data comprising the ordinary data and one of the outliers is labelled by the binary indicator $L_j$, where $L_j=0$ and 1 represent whether the candidate outlier is truly an outlier in the data. Typically, the outliers of the data \emph{is} simplistically modelled with the contamination model with a potential mean shift scale $\delta_j \sim \mathcal{N}(0, \tau^2)$ with a large hyperparameter $\tau \sim 5$ by assuming involvement of another process, so that if the standardised ordinary data follows the normal distribution $\mathcal{N}(0,1)$, the outlier at the right tail follows $\mathcal{N}\left(\left|\delta_j\right|,1\right)$. $\left|\delta_j\right|$ may be interpreted as the effect size. After evaluating the distribution of $h^*$ without an outlier $\mathbb{P}\left(h^*\middle|L=0, \delta=0\right)$ and with an outlier of a particular effect size $\mathbb{P}\left(h^*\middle|L=1, \delta\right)$ (e.g.,~by Monte Carlo simulation), the joint posterior for outlier $j$ can be modelled by the Jeffreys' prior outlier probability $\pi_j \sim \text{Beta}(0.5, 0.5)$, i.e.,~$L_j \sim \text{Bernoulli}(\pi_j)$, without any prior information, given by the Bayes theorem:

\begin{equation} \label{eq:8}
    \mathbb{P}\left(L_j, \pi_j, \delta_j \middle| h^*_{\text{obs},j}\right) \propto \mathbb{P}\left(h^*_{\text{obs},j} \middle| L_j, \delta_j\right) \mathbb{P}\left(L_j \middle| \pi_j\right) \mathbb{P}(\pi_j) \mathbb{P}(\delta_j)
\end{equation}
where $h^*_{\text{obs},j}$ is the value of $h^*$ of the observed data. Marginalising $\delta_j$ to obtain $\mathbb{P}\left(h^*_{\text{obs},j} \middle| L_j\right)$ using

\begin{equation} \label{eq:9}
    \mathbb{P}\left(h^*_{\text{obs},j} \middle| L_j\right) \propto 2 \int_{0}^{\infty} \mathbb{P}\left(h^*_{\text{obs},j} \middle| L_j, \delta_j\right) \mathbb{P}(\delta_j) \, \mathrm{d}\delta_j,
\end{equation}
or Markov chain Monte Carlo so that (\ref{eq:8}) is simplified to

\begin{equation} \label{eq:10}
    \mathbb{P}(L_j, \pi_j | h^*_{\text{obs},j}) \propto \mathbb{P}\left(h^*_{\text{obs},j} \middle| L_j\right) \mathbb{P}\left(L_j \middle| \pi_j\right) P(\pi_j),
\end{equation}
the posterior probability of an outlier (noting $\mathbb{P}\left(L\middle|\pi\right) = \pi^L(1-\pi)^{1-L}$):

\begin{equation} \label{eq:11}
    \mathbb{P}\left(L_j=1 \middle| h^*_{\text{obs},j}\right) = \int_{0}^{1} \frac{\mathbb{P}\left(h^*_{\text{obs},j} \middle| L_j=1\right)\pi_j}{\mathbb{P}\left(h^*_{\text{obs},j} \middle| L_j=1\right)\pi_j + \mathbb{P}\left(h^*_{\text{obs},j} \middle| L_j=0\right)(1-\pi_j)} \cdot \mathbb{P}(\pi_j) \, \mathrm{d}\pi_j.
\end{equation}
Decision rules of outlier inference can then be implemented using the combined posterior for all outliers:

\begin{equation} \label{eq:12}
    \mathbb{P}\left(L_1=\dots=L_{n'}=1 \middle| \mathbf{h}^*_{\text{obs}}\right) = \frac{1}{K} \prod_{j=1}^{n'} \mathbb{P}\left(L_j=1 \middle| h^*_{\text{obs},j}\right)
\end{equation}
where $K$ is the normalization constant for all possible outcomes, i.e.,~all possible combinations of the outliers of the $k$ greatest data points are truly outliers:

\begin{equation} \label{eq:13}
\begin{aligned}
K &= \sum_{k=1}^{n'} \prod_{j=1}^{n'} \mathbb{P}\left(L_j(k) \middle| h^*_{\text{obs},j}\right) \\
  &= \sum_{k=1}^{n'} \prod_{j=1}^{n'} \frac{\mathbb{P}\left(h^*_{\text{obs},j} \middle| L_j(k)\right) \mathbb{P}\left(L_j(k)\right)}{\mathbb{P}(h^*_{\text{obs},j})} \\
  &= \sum_{k=1}^{n'} \prod_{j=1}^{n'} \frac{\mathbb{P}\left(h^*_{\text{obs},j} \middle| L_j(k)\right) \pi_j^{L_j(k)} (1-\pi_j)^{1-L_j(k)}}{\mathbb{P}\left(h^*_{\text{obs},j} \middle| L=1\right)\pi_j + \mathbb{P}\left(h^*_{\text{obs},j} \middle| L=0\right)(1-\pi_j)}
\end{aligned}
\end{equation}
where $L_j(k) = \mathbbm{1}_{[k,n']}(j)$. $\mathbbm{1}_{[k,n']}(j)$ is the indicator function of $j$ in $[k, n']$.

\section{Paired-samples test on $h^*$ for treatment exclusive to extreme cases} \label{sec:paired-test}

In addition to its application in assessing the reliability of outlying observations, the $h^*$ statistic (whether frequentist or Bayesian) may also serve as an indicator of treatment effects that are specific to extreme cases (cf.~individuals in the terminal stages of chronic kidney disease).

Individuals who are not unwell do not respond to treatment. Consequently, in a random population sample, only a small number of cases are likely to exhibit sensitivity to the intervention. These cases would typically fall outside the primary scope of analysis, leading to the conclusion that the treatment is ineffective and thereby not refuting the null hypothesis for the majority of the data, while disregarding those few who are genuinely helped by the intervention.

For instance, suppose a treatment is developed to address loneliness, where the pre-treatment assessment (pretest) identifies the exceptionally lonely people (outliers) who do not exhibit as outliers after the intervention (posttest), indicating that the treatment was effective for those exhibiting the most severe symptoms, while the remainder of participants were unaffected. Let's consider a case study of a generated dataset of size $N=180$, tabulated in Table~\ref{tab:paired-test-raw}. To evaluate this, one could identify the outliers using the procedures described in Section~\ref{sec:h-test-demo}, where every $h^*$ of this group of people is significant (Table~\ref{tab:paired-test-result}). Six outliers were identified prior to intervention (e.g.,~$h^* > 2.85$, $p < .012$ assuming log-normal data), as shown in Table~\ref{tab:paired-h}. The same procedure applied to the posttest data recognised no outliers (i.e.,~no combination of data points gives consistent $p <.05$, as shown in Table~\ref{tab:paired-test-result}). A more specific test is needed to evaluate the intervention effectiveness. The outliers were separated for a close investigation of the $h^*$ statistic, noting its physical meaning as described in Section~\ref{sec:h-formulation} and \ref{sec:choice}:

\begin{table}[h]
    \caption{$h^*$ values of the outliers identified in the pretest of the intervention}
    \label{tab:paired-h}
    \begin{tabular}{ccc}
    Participant No. & $h^*_{\text{pre}}$ & $h^*_{\text{post}}$ \\
    26 & 2.85 & 2.40 \\
    59 & 3.59 & 2.21 \\
    68 & 3.89 & 2.34 \\
    158 & 3.89 & 2.12 \\
    173 & 4.29 & 2.60 \\
    177 & 3.81 & 2.21 \\
    \end{tabular}
\end{table}
If $h^*$ was significant before the treatment but not afterwards for a given participant, this would suggest that the treatment was successful for that extreme case. However, if only some of these six individuals respond to the effectiveness, it becomes difficult to ascertain the overall reliability of the treatment's effectiveness.

Therefore, one may take the six $h^*$ values before and after treatment, regardless of whether they lie on the extreme, and conduct the Wilcoxon signed-rank test (or other applicable paired statistical test, such as paired permutation test on mean) on the $h^*$ values to determine whether the intervention had a statistically significant effect on the extreme cases, specifically in the expected direction (i.e.,~towards reintegration into the normally distributed data). The Wilcoxon signed-rank test offers a paired-sample test for the non-normal $h^*$ data (one may also perform the paired $t$-test if the $h^*$ can be regarded as normal). In this example, the value of the test statistic is 21, corresponding to $p = 0.036$. Therefore, the intervention was significantly effective for the loneliest among people---even when some cases remained extreme (i.e.,~\#173's $h^*$ falls into the rejection region), although less than before.

\section{Generalisation of the $h^*$ test statistic}
\subsection{Weighted distance}
The rms formulation in Definition~\ref{eq:3} may be extended into weighted mean for adaption to specific context. For example, in an ant colony, the discovery and collection of food are organised through a division of labour between a few scout ants and a multitude of forager ants \cite{holldobler1990ants}. Scout ants act as solitary explorers, venturing away from the nest in random, winding paths to search for new food sources or nesting sites. Their journeys are often long and circuitous, covering much more ground than the straight-line distance from the nest, as they investigate unfamiliar territory and maximise the chances of finding resources \cite{10.1146/annurev-ento-010814-020627}.

Scouts walk alone. When a scout ant discovers a valuable food source, it returns to the nest and communicates its find, often by laying down a pheromone trail \cite{Deneubourg1990, Goss1989}. This chemical trail serves as a guide for forager ants, who then leave the nest in groups to exploit the new resource. Unlike scouts, foragers follow these established pheromone trails, walking directly and efficiently between the nest and the food source. Their paths are much more straightforward, so the distance they travel closely matches the actual distance to the food.

This teamwork allows the colony to balance exploration and exploitation: scouts expand the colony’s reach, while foragers efficiently harvest resources. Together, their complementary behaviours ensure the colony’s survival and success in a changing environment.

Some studies used tracking and video analysis to compare the movement patterns of scouts and foragers \cite{10.1146/annurev-ento-010814-020627, Deneubourg1990}. Scouts are at the periphery of the foraging area (mild outliers) or even beyond it (strong outliers), while foragers are normally distributed along established trails and near food sources. In the Argentine ant (\emph{Linepithema humile}), scouts have been observed to leave the main trails and explore further afield, while most workers stick to the trails \cite{holldobler1990ants, Goss1989}.

It follows that to discern a (mild) scout from a forager, sheer distance from the nest may indicate the outlier as compared to the foragers clustered around the nest but that measure is insufficient to discern foragers from the scout they follow to new food sources and nesting sites. Therefore, calculating $h^*$ to tell scout from forager should be based on distance from the majority of foragers weighted by the mileage each individual ant has covered to reach the new site. Mileage alone also does not do it as foragers may circle around in close proximity of the nest. Then, we may attempt

\begin{definition}
The weighted $h^*$ statistics with weights $w_{\alpha\beta}\!\left(\bm{\theta}\right) \ge 0$ associated with members $\alpha$ and $\beta$ and environment parameter $\bm{\theta}$ is

\begin{equation*} 
h^*_\mathrm{w} = \sqrt{\frac{\displaystyle\left.\sum_{k=1}^{n} w_{k,{}^*}\!\left(\bm{\theta}\right) (X_k - X^*)^2 \middle/ \sum_{k=1}^{n}w_{k,{}^*}\!\left(\bm{\theta}\right) \right.}{\displaystyle\left.\sum_{\substack{i>j \\ X_i, X_j \neq X^*}} w_{ij}\!\left(\bm{\theta}\right) (X_i - X_j)^2 \middle/ \sum_{\substack{i>j \\ X_i, X_j \neq X^*}} w_{ij}\!\left(\bm{\theta}\right) \right.}}.
\end{equation*}
\end{definition}
In the above ant example, $w_{k,{}^*}\!\left(\bm{\theta}\right)$ may be a function of $W_{\mathrm{d}}$ that increases with the distance $d$ between member $k$ and the new food source characterised by $\bm{\theta}$. For instance, $W_{\mathrm{d}}$ can simply be an identity function that uses the distance as the weight, or a monotonic convex or concave scale-invariant function for nuisance weighting. A natural choice of $w_{ij}\!\left(\bm{\theta}\right)$ is the geometric mean of $w_{i,{}^*}\!\left(\bm{\theta}\right)$ and $w_{j,{}^*}\!\left(\bm{\theta}\right)$, i.e.,~$w_{ij}\!\left(\bm{\theta}\right) = \sqrt{w_{i,{}^*}\!\left(\bm{\theta}\right) w_{j,{}^*}\!\left(\bm{\theta}\right)}$, as such distance may not be highly regarded if one of the ant pairs is likely to follow the scout ant which does not reflect the normally distributed positions.

Another application concerns psychometric rating scales, such as [0–5] or [1–6], which do not have equal intervals--—that is, the psychological `distance' between adjacent points is not constant. This issue has been explored in several influential studies. Many of these studies suggest that responses on such scales often reflect a log-linear or otherwise non-linear relationship with the underlying psychological construct.

For example, Stevens introduced the idea that the relationship between stimulus intensity and perceived magnitude often follows a power law (a type of log-linear relationship) \cite{1958-04769-001}. While this was originally about sensory magnitudes, the principle has been extended to rating scales in psychology. Thurstone's work showed that subjective judgments (including those made on rating scales) are often distributed normally on an underlying latent continuum, but the mapping from scale points to this continuum is not necessarily linear \cite{1928-00527-001}. Luce worked on probabilistic choice and logit models, underpinning much of the modern understanding that categorical responses (like those on rating scales) often reflect a log-linear relationship with the underlying variable \cite{luce1959individual}. The Rasch model \cite{rasch1960probabilistic}, foundational to Item Response Theory, assumes that the probability of endorsing a particular response category is a logistic (log-linear) function of the difference between person ability and item difficulty. This implies that the intervals between scale points are not equal on the latent trait. These and other empirical studies on non-linearity of rating scales \cite{borg2005modern, 1998-12057-010} show that people interpret and use rating scales in non-linear ways, often compressing or expanding certain parts of the scale. For an overview, consult \cite{Norman2010}, reviewing evidence that Likert-type rating scales are not truly interval while discussing the implications for statistical analysis.

The consensus from psychophysics, measurement theory, and empirical studies is that (1)~there may be subjective variation in how participants perceive rating scales (e.g.,~0--5), and (2)~these scales are often used in ways that reflect a log-linear or otherwise non-linear mapping to the underlying psychological variable. This suggests that the distances used as input for the $h^*$ statistic should be weighted according to their application domain, thereby more accurately reflecting empirical outliers than simply assigning a default value of 1 to all distances.

\subsection{Sensitivity to differences}
The $h^*$ test statistic formalism can be further generalised to various sensitivities by replacing the rms formulation with the H\"older mean \cite{Bullen2003}:

\begin{definition}
The generalised $h^*$ statistics with weights $w_{\alpha\beta}\!\left(\bm{\theta}\right) \ge 0$ associated with members $\alpha$ and $\beta$ and environment parameter $\bm{\theta}$ and the sensitivity parameter $\eta$ is

\begin{equation*} 
h^*_\eta = \left(\frac{\displaystyle\left.\sum_{k=1}^{n} w_{k,{}^*}\!\left(\bm{\theta}\right) (X_k - X^*)^\eta \middle/ \sum_{k=1}^{n} w_{k,{}^*}\!\left(\bm{\theta}\right) \right.}{\displaystyle\left.\sum_{\substack{i>j \\ X_i, X_j \neq X^*}} w_{ij}\!\left(\bm{\theta}\right) (X_i - X_j)^\eta \middle/ \sum_{\substack{i>j \\ X_i, X_j \neq X^*}} w_{ij}\!\left(\bm{\theta}\right) \right.}\right)^{1/\eta}.
\end{equation*}
\end{definition}
The exponent $\eta$ effectively rescales the distance distribution, expanding further for the farther for $\eta>1$, similar to the Hubble's law that describes the recession velocity--distance relationship of interstellar objects in the expanding universe, and vice versa. The higher the value of $\eta$, the higher the sensitivity, implying larger $h^*$ values even for small effect sizes, and hence higher tendency for a data point to be identified as an outlier or more data points to be identified as outliers. 

The sensitivity exponent offers a fine-tuning strategy of optimising the power of the $h^*$-test and hence the probability of Type II errors. The most powerful test, according to the Neyman–-Pearson lemma, would suggest that heavy-tailed distributions or skewed distributions that generally lead to heavier tails in the $h^*$ distribution may fit into the greatest test power with a smaller exponent. This can be explained by starting with the effect of a larger exponent on the test power. A dataset with outliers may be described by a mixture distribution which effectively broadens the $h^*$ distribution with a mode shift. Increasing the exponent rescales the distance in a way that leads to a heavier tail and a shift in the critical value. The degree of shift in the critical value of $h^*$ depends on the shift in the mode and the change in the tail weight. The shift in the mode is supposed to be minor because the shift in the distance of the ordinary data that occupies a central location is relatively small. The shift is greater with lighter tails because the area enclosed per $h^*$ interval at that region is smaller. Note that the mixture distribution, due to the presence of far data points, gives an even more extended tail compared to the ordinary data distribution. Therefore, the broadening effect and hence the increase in the tail weight of the $h^*$ distribution for the mixture distribution is expected to be greater. The overall result of the increase in the exponent is that the extent of broadening and increase in tail weight of the $h^*$ distribution for the mixture distribution is stronger than the ordinary data distribution, where the shift in the critical value of $h^*$ may shift more for an exponent increment at small values but less when the exponent gets sufficiently large, leading to a potential marginal rise-and-fall in and thus optimisation for the test power. Note that the trend of the test power with respect to the exponent depends on the distribution features of the dataset and may also be monotonic.

Besides comparing to a threshold for outlier detection as in Section~\ref{sec:h-test-demo}, the sensitivity parameter also has an effect on inference where actual values of $h^*$ are considered, such as applying the paired permutation test on mean for assessing the significance  of a process---the paired permutation test compares the proportion of the magnitude of permuted mean differences that is greater than the observed mean difference as the $p$-value. Therefore, rescaling the distance distribution may alter the location of the distances relative to their mean and hence the resulting proportion and $p$-value. The significance of the effect of actual $h^*$ modulated by $\eta$ is even more inevitable in Bayesian inference as described in Section~\ref{sec:bayesian}, since the actual probability participates throughout the analysis (until a decision rule is applied that may or may not discard the magnitude of the probability up to categorisation of values).

In subjective inference, agencies rely on their own sensory organs and belief system for signal detection, e.g.,~whether the sea is blue, the tone is friendly. The signal of perception, during information processing in the agency's mental world, is effectively an interference of the sensory organ--encoded (and thus perceptual) information by the belief system that creates the bias of subjectivity \cite{10.1075/lal.12.02ch1}, which can be treated as the observer effect where the epistemics of the agencies, including the beliefs and the processing of the information, entangle with and transform the pre-encoded information and eventually becoming observable or conscious perceptions via some form of measurement \cite{hoorn2024observereffectquantumcase}. In neural terms, sensory processing sensitivity (SPS) accounts for the sensitivity and responsiveness to the environment and social stimuli \cite{10.1002/brb3.242}. Therefore, in modelling subjective inference of the signal as observations of outliers from the noise distribution, individual differences in SPS suggests variable $\eta$ across individuals \cite{10.3389/fpsyg.2022.1010836}, and high SPS people possess higher $\eta$ and see signals (distinct information) more easily out of the null (noise distribution) \cite{1967-02286-000}. In other words, the generalised $h^*$ statistic may serve as a measure of signal detection for agencies with individual differences of sensitivity. In contrast to the objective inference described in the above scenarios, sensitivity in subjective inference is not necessarily correlated to optimised inference. Nonetheless, dynamic values of $\eta$ may offer a description of progressive learning of wise SPS.

\section{The unique outlier}
As demonstrated in the power analysis of the $h^*$-test, the gradual decline (Figure~\ref{fig4}) observed at larger sample sizes (Figure~\ref{fig4} and \ref{fig5}) suggests that the distinctiveness of exceptionality diminishes as the sample size increases, owing to a greater number of cases occupying the previously unrepresented range or `gap' (cf.~(\ref{eq:21})).

Certain things may happen once in a lifetime or an event may transpire but once in a century, which indeed may be deemed unique. However, such an event need not necessarily be an outlier among similar phenomena that unfold in a comparable manner. For instance, consider the rare `great planetary alignment,' which, while infrequent, follows the predictable laws of celestial workings. An individual may be deemed exceptional within the context of their group, yet not singularly unique, for there may exist others who share this exceptional quality, a pertinent example of which is found within the ranks of Mensa International, where numerous members exhibit extraordinary intellectual capabilities. Thus arises the necessity for a metric that distinguishes between uniqueness and the state of being an outlier. This prompts the inquiry: might we discover a person or event, say, muon $g-2$ and $B$ anomalies from dark matter \cite{PhysRevLett.127.061802}, that deviates from its counterparts in a manner that renders it distinct? Furthermore, one must ponder whether it is invariably the same individual or event that emerges as an outlier across various dimensions within each sample. Therefore, a measure that tells outliers from the mean should be accompanied by a metric for being novel and then staying uniquely so.

Creative work often is seen as without limitations because limitations shrink the original design space. True as this may be, limitations also make one evade the clichés. The artists may not allow themselves to use normal language, common imagery, or known patterns. Therefore, in the largest possible sample one can draw, the frequency ($f$) of occurrence of a creative expression should ideally approach~1. On that note, in \cite{10.1145/3574131.3574444}, to be considered `new' or `unique' is to occupy a position of statistical rarity, such that novelty or uniqueness manifests with a frequency of occurrence equating to $f=1$ in relation to the set size of accessible prior information, or the sample of size $n_0$ extracted from the totality of the information universe (of size $N$). Consequently, a unique incidence ($I$) can be expressed as $I = f/n_0$, where $f \ge 1$, ideally maintaining a value of 1 across all samples drawn from $N$. As $f$ exceeds 1, the value of $I$ increases, thereby diminishing the degree of novelty or uniqueness achieved.

Frequently, we observe that $I = 1/n_0 \neq f/N$ \cite{10.1145/3574131.3574444}. As the sample size $n_0$ enlarges, the probability that $f > 1$ also rises. For the same observer, $I = 1/10$ is less novel or unique than $I = 1/100,000$ and so the value of $I$ asymptotically approaches zero, though it should never quite reach it.

If novelty or being unique is a requisite, the aspirational state is $I = 1/N$ \cite{10.1145/3574131.3574444}. Theoretically, $N$ remains unknowable, as no observer possesses complete knowledge of the information universe's expanse \cite{10.1075/lal.12.02ch1}. Fortunately, the fulfilment of this goal is often permitted to be partial: a solution that is sufficiently novel may be satisfying already. Therefore, one may assert that the demand for novelty of $I$ should diminish across a series of samples. In a slightly adapted version of \cite{10.1145/3574131.3574444}:

\begin{equation} \label{eq:*}
    I = \frac{1}{n_0} \ge \frac{f}{\sum_{i=0} n_i},
\end{equation}
where $\Sigma_i n_i \le N$ and $n_i$ ($i \ge 1$) represents the additional sample drawn beyond the initial sample (size $n_0$) in the sum of samples.

\cite{10.1145/3574131.3574444} points out that in the following sequence, the first line (with a tick) signifies the discovery of the most unique incidence, the second (also with a tick) remains acceptable, whereas the third (with a cross) indicates a biased, myopic sample in $n_1$, disregarding `compromising' information.

\begin{tabular}{llllll}
    \textbf{$n_1$} & & \textbf{$n_2$} & & \textbf{$n_3$} & \\
           $1/10$ & $>$ & $1/100$ & $>$ & $1/1,000$ &  $\checkmark$ \\
           $1/10$ & $=$ & $10/100$ & $=$ & $100/1,000$ & $\checkmark$ \\
           $1/10$ & $<$ & $20/100$ & $<$ & $500/1,000$ & $\times$ \\
           \multicolumn{1}{c}{$\uparrow$}  \\
\end{tabular}

Too small a sample.
\vspace{1em}

Consider, for Observer~1, that $I = 1/1 > 1/10^5$, where $N = 10^5$ denotes the total number of known rare-earth element-based superconducting magnets within that individual's cognitive repertoire or within the relevant scientific knowledge base. In this context, the discovery of a superconducting magnet utilising a novel yttrium-barium-copper-oxide (YBCO) configuration is perceived as `highly innovative.' In contrast, for Observer 2, $I = 1/5,000 < 5,000/10^5$, where 5,000 represents the entirety of known YBCO-based superconducting magnets, the novelty of the configuration is substantially diminished, and the finding is regarded as `relatively commonplace.' However, if one were to integrate all known YBCO-based superconducting magnets into a single, large-scale quantum computing array, $I = 1/1 > 1/10^9$, where $N = 10^9$ encompasses all known quantum computing arrays worldwide, even Observer~2 would be compelled to acknowledge the configuration as a `genuinely novel contribution,' the materialisation of a new idea.

If we were to make a conceptual classification of cases based on the $h^*$ statistic and the $I$-index, a different combination of high or low values for these metrics corresponds to the following interpretation:

\begin{itemize}[leftmargin=*]
    \item High $h^*$, high $I$. This quadrant denotes exceptionality that is recurring. Cases falling into this category are statistically outlying (high $h^*$) yet frequently observed (high $I$), suggesting a pattern of repeated outstanding performance or phenomena that is not unique.
    \item High $h^*$, low $I$. This cell represents genius that is unique. Here, the case is statistically exceptional (high $h^*$) as well as rarely observed, preferably only once (low $I$), indicating a singular, extraordinary occurrence or individual.
    \item Low $h^*$, high $I$. This combination describes above-average or even normal cases that are commonly found. The cases are not statistically exceptional (low $h^*$) and are frequently encountered (high $I$), reflecting a high prevalence of moderately or even quasi superior instances.
    \item Low $h^*$, low $I$. This quadrant refers to above-average to normal cases that are rarely found (e.g.,~among high achievers). These cases are neither statistically exceptional (low $h^*$) nor commonly observed (low $I$), suggesting infrequent occurrences of moderate merit.
\end{itemize}

Overall, the aforementioned would be a framework for interpreting the intersection of statistical exceptionality ($h^*$) and uniqueness of an incidence ($I$), facilitating nuanced distinctions between different types of notable cases.

\section{Discussion and Conclusions}
The present study has introduced and evaluated the $h^*$ ($h$-star) test statistic as a novel parametric and frequentist approach (with Bayesian potential) for the assessment of global outliers within one-dimensional homogeneous datasets, without necessary recourse to the assumption of normality. In so doing, we have sought to address a critical lacuna in the extant statistical literature, wherein conventional outlier detection methods, such as Grubbs' test and Dixon's $Q$, are not only predicated upon the presumption of Gaussianity, but also tend to conceptualise outliers as aberrations to be excised in the pursuit of statistical, we say nonsensical, `purity.' By contrast, the $h^*$ statistic is expressly designed to foreground the interpretative and substantive value of outliers, treating them as phenomena meriting investigation in their own right.

Our empirical demonstration, utilising data from mood intervention studies with social robots, has elucidated the practical utility of $h^*$ in distinguishing between stably extraordinary deviations and those values that merely appear extreme under traditional criteria. The $h^*$ statistic, by quantifying the extremity of a candidate outlier relative to its group through a ratio of rms pairwise distances, provides a measure of statistical significance and confidence that is analogous to the role of Student's $t$ in the comparison of means. Notably, the $h^*$-test is robust to deviations from normality and is invariant under linear transformations, thereby extending its applicability to a wide array of empirical contexts.

The power analysis of the $h^*$-test has revealed that its sensitivity to outliers increases with both effect size and sample size, yet exhibits a gentle attenuation at larger sample sizes. This phenomenon is attributable to the progressive occupation of the distributional `gap' by additional cases, thereby diminishing the distinctiveness of exceptionality as the sample becomes more representative of the underlying population. This observation underscores a fundamental epistemological point: the uniqueness of an outlier is not an intrinsic property, but is contingent upon the scope and granularity of the sampled data. As the sample size increases, the probability of observing extreme values rises, and the singularity of any given outlier correspondingly wanes.

Furthermore, the $h^*$ framework facilitates a nuanced distinction between statistical exceptionality and uniqueness, as articulated through the intersection of the $h^*$ statistic and the $I$-index of novelty. This dual-metric approach enables the differentiation of recurring exceptionality (high $h^*$, high $I$), singular genius (high $h^*$, low $I$), common above-average cases (low $h^*$, high $I$), and rare but ordinary instances (low $h^*$, low $I$). Such a taxonomy is of particular relevance in domains where the identification of both extraordinary and unique cases is of substantive interest, for example in talent identification, clinical diagnostics, or the detection of rare events in quantum physics or cyber-security.

The implications of the $h^*$-test extend beyond mere methodological innovation. By challenging the hegemony of normality and the reflexive excision of outliers, the $h^*$ statistic invites a reorientation of statistical practice towards a more inclusive and interpretatively rich engagement with data. Outliers, rather than being dismissed as statistical artefacts, are repositioned as potential harbingers of novel phenomena, rare pathologies, or exceptional talent. This perspective is consonant with the epistemic virtues of scientific inquiry, which prizes the anomalous as a potential source of theoretical advancement.

Moreover, the $h^*$ statistic offers a principled means of evaluating the efficacy of interventions targeted at extreme cases, as demonstrated through the application of paired-sample tests on $h^*$ values for pre- and post-interventions. This approach enables the quantification of treatment effects specifically among those individuals who are most deviant from the norm, thereby addressing a recurrent limitation of conventional inferential statistics, which often obscures such effects within aggregate analyses.

The generalised $h^*$ test statistic allows for controlling the sensitivity for outliers via the sensitivity exponent $\eta$ and a comprehensive control mechanism over the test statistic's behaviour via the adaptive weights that controls the influence of individual differences depending on the individual or pair-wise characteristics and the environment, extending simple distance measures to adapt to domain context. The generalised $h^*$-test can adapt to various distribution features such as skewness and tail weight for a potentially optimised power that fit the contextual requisite. It also provides refinement to the contextual interpretation of how distinct a data point is for specific measures and analyses. Besides objective inference enhancement, it may serve as a customisable measure for signal detection of subjective inference with various sensitivities. These functionalities make the $h^*$-test stand out from other comparable tests, enabling contextual, realistic and nuance outlier analyses. 

In conclusion, the $h^*$-test statistic constitutes a robust and versatile addition to the statistical toolkit for outlier analysis. Its capacity to operate independently of distributional assumptions, to provide interpretable measures of exceptionality, and to facilitate the nuanced classification of outliers according to both statistical and substantive criteria, renders it particularly valuable in the analysis of complex, real-world datasets. Future research may profitably extend the $h^*$ framework to multivariate contexts, explore its integration with Bayesian inferential paradigms, and further elaborate its implications for the philosophy of statistical practice.

In so doing, we may move towards a more comprehensive and inclusive understanding of the extraordinary, the unique, and the anomalous within empirical science. The $h^*$ test statistic and associated $I$-index constitute a defence of those artists, designers, scientists, and engineers whose work is often regarded as excessively eccentric or unconventional by prevailing standards of habit and taste.

\begin{appendix}

\section{Proof of the $h^*$ expression (Lemma~\ref{lem:h-formula-alt})} \label{sec:h-eq-proof}
The numerator is straightforward:

\label{eq:14}
\begin{align}
    \sum_{k=1}^{n} (X_k - X^*)^2 &= \sum_{k=1}^{n} X_k^2 - \sum_{k=1}^{n} 2X^*X_k + \sum_{k=1}^{n} X^{*2} \nonumber \\
                                 &= \sum_{k=1}^{n} X_k^2 - 2X^* \sum_{k=1}^{n} X_k + nX^{*2}
\end{align}
For the denominator, noting $\sum_{i>j} (X_i - X_j)^2 = n^2\mathbb{V}(X)$, $\sum_{\substack{i>j \\ X_i,X_j \neq X^*}} (X_i - X_j)^2$ effectively drops $X^*$ from the set of random variables, thus

\label{eq:15}
\begin{align}
\sum_{\substack{i>j \\ X_i,X_j \neq X^*}} (X_i - X_j)^2 &= (n-1)\left(\sum_{X_k \neq X^*} X_k^2\right) - \left(\sum_{X_k \neq X^*} X_k\right)^2  \nonumber \\
&= (n-1)\left[\left(\sum_{k=1}^{n} X_k^2\right) - X^{*2}\right] - \left[\left(\sum_{k=1}^{n} X_k\right) - X^*\right]^2. 
\end{align}
\pagebreak

\section{Proof of (\ref{eq:5})} \label{sec:h-full-proof}
Since $h^*$ is a function of all $X_k$, it is desirable to transform the random variables $\mathbf{X}=(X_k)$ such that $h^*$ can be expressed with fewer dimensions. There are $n!$ permutations of $X_k$ that gives the same $h^*$ (the transformation is many-to-one). For the sake of simplicity, we arrange $X_k$ and define the ordered statistics $X_{(k)}$ such that $X_{(i)} \le X_{(j)}$ iff $i \le j$. Note $X_{(n)} = \max\{X_k\}$. Under this transformation, the joint pdf of the ordered statistics is

\begin{equation} \label{eq:16}
f_{X_{(\cdot)}}\!\left(\mathbf{x}_{(\cdot)}\right) = n! f_{X_{\left(\cdot\right)}}\!(\mathbf{x}) = n! \prod_{k=1}^n f_X(x_k) \cdot \mathbbm{1}\!\left(x_1 \le x_2 \le ... \le x_n\right). 
\end{equation}
Note that the support of $X_{(k)}$ is the same as $X_k$. Now, define $U_1 = X_{(n)}$, $U_{k+1} = X_{(n)} - X_{(k)} \ge 0$ ($1 \le k \le n-1$). We obtain the inverse

\begin{equation} \label{eq:17}
X_{(k)} = \begin{cases} U_1, & k=n \\ U_1 - U_{n-k+1}, & 1 \le k \le n-1 \end{cases}
\end{equation}
and the corresponding joint pdf for $\mathbf{U}=\mathbf{u}$

\begin{equation} \label{eq:18}
f_{U}\!\left(\mathbf{u}\right) = f_{X_{(\cdot)}}\!\left(\mathbf{x}_{(\cdot)}\!\left(\mathbf{u}\right)\right) \left| \frac{\partial \mathbf{x}_{(\cdot)}}{\partial \mathbf{u}} \right| = n! f_X(u_1) \prod_{k=2}^{n} f_X(u_1 - u_k) \cdot \mathbbm{1}\!\left(u_2 \ge u_3 \ge ... \ge u_n \ge 0\right)
\end{equation}
with the absolute determinant of the Jacobian $\left| \frac{\partial \mathbf{x}_{(\cdot)}}{\partial \mathbf{u}} \right| = 1$ because the transformation is essentially a shift (with mirroring) without rescaling.
Then, the numerator becomes

\begin{equation} \label{eq:19}
\sum_{k=1}^{n} \left[X_k - \max\left\{\mathbf{X}\right\}\right]^2 = \sum_{k=1}^{n-1} \left(X_{(n)} - X_{(k+1)}\right)^2 = \sum_{k=2}^{n} U_k^2, 
\end{equation}
and the denominator

\begin{equation} \label{eq:20}
    \begin{aligned}[b]
    & (n-1) \left[ \left(\sum_{k=1}^n X_k^2\right) - X^{*2} \right] - \left[ \left(\sum_{k=1}^n X_k\right) - X^* \right]^2 \\
    & = (n-1)(n-2)\mathbb{V}_{n-1}\!\left(X_{(\cdot)}\right) \\
    & = (n-1)(n-2)\mathbb{V}_{n-1}\!\left(U_1 - U_{n-\cdotp+1}\right) \\
    & = (n-1)(n-2)\mathbb{V}_{n-1}\!\left(U\right) \\
    & = (n-1) \sum_{k=2}^n U_k^2 - \left( \sum_{k=2}^n U_k \right)^2
    \end{aligned}
\end{equation}
where $\mathbb{V}_{n-1}$ denotes the sample variance of the first (last) $(n-1)$ $X_{(k)}$ ($U_k$). Therefore, Lemma~\ref{lem:h-formula-alt}, (\ref{eq:19}) and (\ref{eq:20}) imply

\begin{equation} \label{eq:21}
    \tilde{h}^* = \sqrt{\frac{\displaystyle\sum_{k=2}^n U_k^2}{(n-1)\displaystyle\sum_{k=2}^n U_k^2 - \left(\displaystyle\sum_{k=2}^n U_k\right)^2}} = \frac{1}{\sqrt{(n-1) - \displaystyle\frac{Q^2}{R^2}}}
\end{equation}
where $Q=\sum_{k=2}^n U_k$ and $R^2 = \sum_{k=2}^n U_k^2$. Now, we have simplified the statistic to a dependence on the ratio between the sum of absolute differences of $X_k$ ($Q$) and the square root of the sum of squares of the absolute difference of $X_k$ ($R$), together with the sample size $n$. The transformation essentially projects the essential information for $h^*$ from the $n$-dimensional sample space to a single dimension, leaving the remaining $n-1$ dimensions as a nuisance. The nuisance variables (all but $h^*$) contain all other necessary information for the sample (e.g.,~the encoded absolute values, possibly together with $h^*$) that is irrelevant to the current scope of interest. Note that the Cauchy-Schwarz inequality guarantees $1 \le Q^2/R^2 \le n-1$ for non-negative $U_k$'s, and thus that all combinations of $X_k$ can be accommodated.

Next, noting that $Q$ and $R$ contain $\left(n-1\right)$ $U_k$'s without $U_1$, we proceed with a transformation from $(U_2, \dots, U_n)$ to another space with $Q/R$ by defining the vector $\mathbf{V} = (U_2 \ \dots \ U_n)^\intercal = R\bm{\hat{\omega}}$, where $\bm{\hat{\omega}}=(\omega_k) \in S^{n-2}$ is a vector in the unit $(n-2)$-sphere as the direction of $\mathbf{V}$, and naturally $R=\left|\mathbf{V}\right|$. Then, $Q$ can be written as the sum of elements of $\mathbf{V}$, i.e.,~$Q = \sum_{k=1}^{n-1} U_{k+1} = R \sum_{k=1}^{n-1} \omega_k$. Now, we express $\sum \omega_k$ with the direction cosine $\cos\varTheta_1$ as the projection of $\bm{\hat{\omega}}$ to $\mathbf{\hat{n}} = \frac{1}{\sqrt{n-1}}(1 \ ... \ 1)^\intercal \in S^{n-2}$ such that

\begin{equation} \label{eq:22}
    \cos\varTheta_1 = \bm{\hat{\omega}} \cdot \mathbf{\hat{n}} = \frac{1}{\sqrt{n-1}}\sum_{k=1}^{n-1} \omega_k = \frac{1}{\sqrt{n-1}}\frac{Q}{R} \in \left[\frac{1}{\sqrt{n-1}}, 1\right].
\end{equation}
Then, we have

\begin{equation} \label{eq:23}
    \tilde{h}^* = \frac{1}{\sqrt{(n-1)-(n-1)\cos^2\varTheta_1}} = \frac{\csc\varTheta_1}{\sqrt{n-1}}.
\end{equation}
We express $\mathbf{V}$ in the hyperspherical coordinates $(R, \varTheta_1\! (\tilde{h}^*), ..., \varTheta_{n-2})$ in the positive hyper-octant, given $U_k \ge 0$ ($k \ge 2$), i.e.,~$R \ge 0$, $\varTheta_k \in [0, \pi/2]$ (constraints of ordering to be imposed). With

\begin{equation} \label{eq:24}
    \sin\varTheta_1 = \frac{1}{\sqrt{n-1}\tilde{h}^{*}},
\end{equation}

\begin{equation} \label{eq:25}
    \cos\varTheta_1 = \frac{\sqrt{(n-1)\tilde{h}^{*2}-1}}{\sqrt{n-1}\tilde{h}^*},
\end{equation}

\begin{equation} \label{eq:26}
    \left|\frac{\partial \varTheta_1}{\partial \tilde{h}^*}\right| = \frac{1}{\tilde{h}^*\sqrt{(n-1)\tilde{h}^{*2}-1}},
\end{equation}

\begin{align} \label{eq:27}
U_k &= \begin{cases}
R \cos\varTheta_1, & k=2 \\
R \left( \displaystyle\prod_{j=2}^{k-2} \sin\varTheta_j \right) \cos\varTheta_{k-1}, & 3 \le k \le n-1 \\
R \displaystyle\prod_{j=2}^{n-1} \sin\varTheta_j, & k=n
\end{cases} \nonumber \\
&= \begin{cases}
\displaystyle\frac{R}{\tilde{h}^*} \sqrt{\frac{(n-1)\tilde{h}^{*2}-1}{n-1}}, & k=2 \\
\displaystyle{\frac{R}{\sqrt{n-1}\tilde{h}^*} \left(\prod_{j=2}^{k-2} \sin\varTheta_j \right) \cos\varTheta_{k-1}}, & 3 \le k \le n-1 \\
\displaystyle{\frac{R}{\sqrt{n-1}\tilde{h}^*} \prod_{j=2}^{n-2} \sin\varTheta_j}, & k=n
\end{cases},
\end{align}
and the inequality $U_2 \ge U_3 \ge ... \ge U_n$ rewritten as $\left(\bigwedge_{k=1}^{n-3} \cot\varTheta_k \ge \cos\varTheta_{k+1}\right) \wedge (\cot\varTheta_{n-2} \ge 1)$, the joint pdf comprises of the $n$ dimensions of coordinates $(u_1, R, \varTheta_1\! \left(\tilde{h}^*\right), ..., \varTheta_{n-2})$ and is given by

\begin{equation} \label{eq:28}
\begin{aligned}
& \quad f_{\mathrm{hs}}\!\left(u_1, R, \tilde{h}^*, \varTheta_2, ..., \varTheta_{n-2}\right) \nonumber \\
&= f_{U}\!\left(\mathbf{u}\!\left(u_1, R, \tilde{h}^*, \varTheta_2, ..., \varTheta_{n-2}\right)\right) \left|\frac{\partial \varTheta_1}{\partial \tilde{h}^*}\right| \cdot R^{n-2} \prod_{k=1}^{n-3} \sin^{n-2-k}\varTheta_{k} \nonumber \\
&= n! f_X\!\left(u_1\right) f_X\!\left(u_1 - \frac{R}{\tilde{h}^*}\sqrt{\frac{(n-1)\tilde{h}^{*2}-1}{n-1}}\right) \nonumber \\
& \times \prod_{k=3}^{n-1} f_X\!\left(u_1 - \frac{R}{\sqrt{n-1}\tilde{h}^*} \left[\prod_{j=2}^{k-2} \sin\varTheta_j\right] \cos\varTheta_{k-1}\right) \nonumber \\
& \times f_X\!\left(u_1 - \frac{R}{\sqrt{n-1}\tilde{h}^*} \prod_{j=2}^{n-2} \sin\varTheta_j \right) \nonumber \\
& \times \mathbbm{1}\!\Biggl(\left( \tilde{h}^* \ge \frac{1}{\sqrt{n-2}} \right) \wedge \left(\sqrt{(n-1)\tilde{h}^{*2}-1} \ge \cos\varTheta_2 \right) \wedge \left(\bigwedge_{k=2}^{n-3} \cot\varTheta_k \ge \cos\varTheta_{k+1}\right) \nonumber \\
& \qquad \wedge (\cot\varTheta_{n-2} \ge 1) \Biggr) \nonumber \\
& \times \frac{R^{n-2}}{\left(\sqrt{n-1}\tilde{h}^*\right)^{n-3}} \prod_{k=2}^{n-3} \sin^{n-k-2}\varTheta_k \cdot \frac{1}{\tilde{h}^*\sqrt{(n-1)\tilde{h}^{*2}-1}},
\end{aligned}
\end{equation}
which is (\ref{eq:4}). Note that $\cot\varTheta_{n-2} \ge 1 \Rightarrow \varTheta_{n-2} \in \left(0,\pi/4\right]$. $\partial\varTheta_1/\partial\tilde{h}^*$ results from the chain rule. The range of $Q/R$ and the expression of $\partial\varTheta_1/\partial\tilde{h}^*$ define the support of $\tilde{h}^* \in [\frac{1}{\sqrt{n-2}}, \infty)$ and hence $h^* \in [\frac{1}{\sqrt{2}}, \infty)$ (Lemma~\ref{lem:hRange}). The marginal distribution of $\tilde{h}^*$ is given by the integral where the limits of $u_1, R$ and $\varTheta_{n-2}$ do not depend on other variables, i.e.,~(\ref{eq:4}). The proof of Theorem~\ref{thm:h-pdf} is complete.

\section{Proof of invariance of $h^*$ and $\tilde{h}^*$ under linear transformations (Lemma~\ref{lem:h-invariance})}\label{sec:h-invariance}
$h^*$ and $\tilde{h}^*$ differ with a scalar only. Taking standardisation ($Z=(X-\mu)/\sigma$) as an example, by expanding

\begin{equation} \label{eq:30}
\begin{aligned}
\left(\sum_{k=1}^n Z_k\right)^2 &= n^2 \mathbb{E}\left(\frac{X-\mu}{\sigma}\right)^2 = \frac{n^2}{\sigma^2}\left[\mathbb{E}(X)-\mu\right]^2 = \frac{n^2}{\sigma^2}\left[\mathbb{E}(X^2)-2\mu \mathbb{E}(X)+\mu^2\right] \\
\end{aligned}
\end{equation}

and
\begin{equation} \label{eq:31}
\begin{aligned}
\sum_{k=1}^n Z_k^2 &= n \mathbb{E}\left[\left(\frac{X-\mu}{\sigma}\right)^2\right] = \frac{n}{\sigma^2}\left[\mathbb{E}(X^2)-2\mu \mathbb{E}(X)+\mu^2\right],
\end{aligned}
\end{equation}
we can deduce with (\ref{eq:20})

\begin{equation*} \label{eq:32}
\begin{aligned}[b]
\tilde{h}^*\{Z\}^2 &= \frac{\displaystyle{\frac{n}{\sigma^2}\left[\mathbb{E}(X^2)-2\mu \mathbb{E}(X)+\mu^2\right] - 2\left(\frac{X^*-\mu}{\sigma}\right) \frac{n}{\sigma}\left[\mathbb{E}(X)-\mu\right] + n\left(\frac{X^*-\mu}{\sigma}\right)^2}}{\displaystyle\mathbb{V}_{n-1}\!\left(\frac{X_{(-)}-\mu}{\sigma}\right)} \\
&= n\frac{[\mathbb{E}(X^2)-2\mu \mathbb{E}(X)+\mu^2] - 2(X^*-\mu)[\mathbb{E}(X)-\mu] + (X^*-\mu)^2}{\mathbb{V}_{n-1}(X_{(\cdot)})}\\
&= \frac{n[\mathbb{E}(X^2)-2X^*\mathbb{E}(X)+X^{*2}]}{\mathbb{V}_{n-1}(X_{(\cdot)})}\\
&= \frac{\displaystyle{n\sum_{k=1}^n X_k^2 - 2X^* \sum_{k=1}^n X_k + nX^{*2}}}{\displaystyle{(n-1)\left[\left(\sum_{k=1}^n X_k^2\right)-X^{*2}\right] - \left[\left(\sum_{k=1}^n X_k\right)-X^*\right]^2}} \\
&= \tilde{h}^*\{X\}^2.
\end{aligned}
\end{equation*}

\pagebreak

\section{Numerical values of critical values of $h^*$}
\begin{table}[htbp]
    \centering
    \caption{Critical values of $h^*$ ($h^*_{\mathrm{c}}$) with $\nu$ degrees of freedom from $n$ i.i.d. random variables of normal distribution, i.e.,~$X_k \sim \mathcal{N}\left(0,1\right)$}
    \label{tab:normal-table}
    \begin{tabular}{*{11}c}
      \multirow{4}{*}{\diagbox[innerwidth=1.2em,height=3em,linewidth=.5pt,dir=SW,innerleftsep=-6pt,innerrightsep=0pt]{$\left.n\,\middle|\,\nu\right.$}{$h^*_{\mathrm{c}}$}} & \multicolumn{10}{c}{\textbf{Single-tail Signifiance Level}} \\
       & 0.40 & 0.30 & 0.20 & 0.15 & 0.1 & 0.05 & 0.02 & 0.01 & 0.002 & 0.001 \\
       & \multicolumn{10}{c}{\textbf{Confidence Level}} \\
       & 60\% & 70\% & 80\% & 85\% & 90\% & 95\% & 98\% & 99\% & 99.8\% & 99.9\% \\
      \hline 
    4 \textbar\ 2 & 1.6442 & 1.9506 & 2.4533 & 2.8698 & 3.5598 & 5.0985 & 8.1204 & 11.5144 & 25.7621 & 36.3713 \\
    5 \textbar\ 3 & 1.5922 & 1.8114 & 2.1488 & 2.4116 & 2.8198 & 3.6413 & 5.0345 & 6.3944 & 11.0270 & 13.9160 \\
    6 \textbar\ 4 & 1.5874 & 1.7683 & 2.0381 & 2.2416 & 2.5479 & 3.1328 & 4.0536 & 4.8883 & 7.4422 & 8.8969 \\
    7 \textbar\ 5 & 1.5957 & 1.7547 & 1.9876 & 2.1599 & 2.4139 & 2.8839 & 3.5901 & 4.2020 & 5.9534 & 6.8873 \\
    8 \textbar\ 6 & 1.6078 & 1.7529 & 1.9622 & 2.1150 & 2.3379 & 2.7415 & 3.3286 & 3.8226 & 5.1768 & 5.8686 \\
    9 \textbar\ 7 & 1.6200 & 1.7563 & 1.9492 & 2.0888 & 2.2904 & 2.6501 & 3.1622 & 3.5832 & 4.6973 & 5.2508 \\
    10 \textbar\ 8 & 1.6321 & 1.7621 & 1.9433 & 2.0731 & 2.2593 & 2.5880 & 3.0483 & 3.4197 & 4.3805 & 4.8445 \\
    11 \textbar\ 9 & 1.6438 & 1.7688 & 1.9414 & 2.0640 & 2.2386 & 2.5446 & 2.9673 & 3.3042 & 4.1598 & 4.5667 \\
    12 \textbar\ 10 & 1.6551 & 1.7762 & 1.9419 & 2.0587 & 2.2242 & 2.5123 & 2.9062 & 3.2171 & 3.9923 & 4.3563 \\
    13 \textbar\ 11 & 1.6660 & 1.7836 & 1.9438 & 2.0560 & 2.2144 & 2.4883 & 2.8593 & 3.1493 & 3.8653 & 4.1950 \\
    14 \textbar\ 12 & 1.6762 & 1.7911 & 1.9466 & 2.0552 & 2.2077 & 2.4699 & 2.8223 & 3.0961 & 3.7647 & 4.0704 \\
    15 \textbar\ 13 & 1.6861 & 1.7986 & 1.9502 & 2.0557 & 2.2031 & 2.4553 & 2.7930 & 3.0538 & 3.6852 & 3.9712 \\
    16 \textbar\ 14 & 1.6956 & 1.8060 & 1.9543 & 2.0570 & 2.2001 & 2.4443 & 2.7690 & 3.0181 & 3.6176 & 3.8866 \\
    17 \textbar\ 15 & 1.7047 & 1.8133 & 1.9587 & 2.0590 & 2.1986 & 2.4356 & 2.7497 & 2.9893 & 3.5635 & 3.8185 \\
    18 \textbar\ 16 & 1.7134 & 1.8203 & 1.9631 & 2.0613 & 2.1977 & 2.4285 & 2.7334 & 2.9654 & 3.5165 & 3.7605 \\
    19 \textbar\ 17 & 1.7218 & 1.8272 & 1.9677 & 2.0642 & 2.1979 & 2.4233 & 2.7199 & 2.9448 & 3.4746 & 3.7085 \\
    20 \textbar\ 18 & 1.7298 & 1.8340 & 1.9724 & 2.0673 & 2.1984 & 2.4191 & 2.7085 & 2.9269 & 3.4406 & 3.6675 \\
    21 \textbar\ 19 & 1.7374 & 1.8404 & 1.9769 & 2.0703 & 2.1992 & 2.4156 & 2.6986 & 2.9121 & 3.4118 & 3.6300 \\
    22 \textbar\ 20 & 1.7448 & 1.8467 & 1.9815 & 2.0736 & 2.2006 & 2.4131 & 2.6903 & 2.8993 & 3.3852 & 3.5968 \\
    23 \textbar\ 21 & 1.7520 & 1.8528 & 1.9861 & 2.0770 & 2.2021 & 2.4111 & 2.6830 & 2.8873 & 3.3618 & 3.5682 \\
    24 \textbar\ 22 & 1.7590 & 1.8589 & 1.9908 & 2.0806 & 2.2041 & 2.4103 & 2.6777 & 2.8780 & 3.3426 & 3.5445 \\
    25 \textbar\ 23 & 1.7656 & 1.8647 & 1.9952 & 2.0840 & 2.2060 & 2.4091 & 2.6720 & 2.8684 & 3.3230 & 3.5206 \\
    26 \textbar\ 24 & 1.7721 & 1.8703 & 1.9997 & 2.0876 & 2.2082 & 2.4088 & 2.6676 & 2.8613 & 3.3088 & 3.5008 \\
    27 \textbar\ 25 & 1.7784 & 1.8759 & 2.0042 & 2.0913 & 2.2105 & 2.4088 & 2.6643 & 2.8548 & 3.2936 & 3.4817 \\
    28 \textbar\ 26 & 1.7843 & 1.8811 & 2.0083 & 2.0946 & 2.2128 & 2.4086 & 2.6609 & 2.8485 & 3.2797 & 3.4640 \\
    29 \textbar\ 27 & 1.7903 & 1.8864 & 2.0127 & 2.0981 & 2.2151 & 2.4089 & 2.6581 & 2.8434 & 3.2672 & 3.4486 \\
    30 \textbar\ 28 & 1.7960 & 1.8916 & 2.0170 & 2.1018 & 2.2177 & 2.4096 & 2.6558 & 2.8386 & 3.2573 & 3.4363 \\
    31 \textbar\ 29 & 1.8016 & 1.8966 & 2.0211 & 2.1053 & 2.2203 & 2.4104 & 2.6540 & 2.8348 & 3.2476 & 3.4232 \\
    32 \textbar\ 30 & 1.8070 & 1.9014 & 2.0251 & 2.1087 & 2.2228 & 2.4110 & 2.6524 & 2.8312 & 3.2390 & 3.4127 \\
    42 \textbar\ 40 & 1.8540 & 1.9443 & 2.0618 & 2.1408 & 2.2479 & 2.4238 & 2.6469 & 2.8112 & 3.1815 & 3.3368 \\
    52 \textbar\ 50 & 1.8920 & 1.9796 & 2.0930 & 2.1690 & 2.2718 & 2.4399 & 2.6523 & 2.8076 & 3.1562 & 3.3020 \\
    62 \textbar\ 60 & 1.9237 & 2.0091 & 2.1196 & 2.1935 & 2.2933 & 2.4561 & 2.6608 & 2.8098 & 3.1423 & 3.2813 \\
    72 \textbar\ 70 & 1.9509 & 2.0349 & 2.1432 & 2.2155 & 2.3130 & 2.4715 & 2.6705 & 2.8150 & 3.1363 & 3.2702 \\
    82 \textbar\ 80 & 1.9748 & 2.0574 & 2.1639 & 2.2350 & 2.3307 & 2.4861 & 2.6807 & 2.8220 & 3.1348 & 3.2645 \\
    92 \textbar\ 90 & 1.9960 & 2.0777 & 2.1826 & 2.2525 & 2.3466 & 2.4994 & 2.6908 & 2.8297 & 3.1366 & 3.2641 \\
    102 \textbar\ 100 & 2.0151 & 2.0958 & 2.1996 & 2.2687 & 2.3618 & 2.5122 & 2.7008 & 2.8371 & 3.1380 & 3.2618 \\
    202 \textbar\ 200 & 2.1428 & 2.2184 & 2.3154 & 2.3795 & 2.4659 & 2.6051 & 2.7782 & 2.9034 & 3.1789 & 3.2926 \\
    302 \textbar\ 300 & 2.2183 & 2.2915 & 2.3850 & 2.4470 & 2.5304 & 2.6638 & 2.8317 & 2.9521 & 3.2155 & 3.3240 \\
    402 \textbar\ 400 & 2.2717 & 2.3434 & 2.4349 & 2.4958 & 2.5769 & 2.7081 & 2.8705 & 2.9885 & 3.2468 & 3.3529 \\
    502 \textbar\ 500 & 2.3131 & 2.3837 & 2.4741 & 2.5333 & 2.6134 & 2.7430 & 2.9037 & 3.0191 & 3.2725 & 3.3761 \\
    602 \textbar\ 600 & 2.3467 & 2.4165 & 2.5054 & 2.5645 & 2.6430 & 2.7702 & 2.9292 & 3.0431 & 3.2940 & 3.3976 \\
    702 \textbar\ 700 & 2.3751 & 2.4442 & 2.5321 & 2.5909 & 2.6680 & 2.7948 & 2.9504 & 3.0644 & 3.3128 & 3.4138 \\
    802 \textbar\ 800 & 2.3989 & 2.4680 & 2.5548 & 2.6132 & 2.6907 & 2.8161 & 2.9716 & 3.0838 & 3.3280 & 3.4280 \\
    902 \textbar\ 900 & 2.4211 & 2.4884 & 2.5759 & 2.6332 & 2.7091 & 2.8339 & 2.9881 & 3.1004 & 3.3459 & 3.4465 \\
    1002 \textbar\ 1000 & 2.4405 & 2.5076 & 2.5945 & 2.6516 & 2.7275 & 2.8494 & 3.0042 & 3.1148 & 3.3565 & 3.4575
    \end{tabular}
\end{table}
\pagebreak

\begin{table}[htbp]
    \centering
    \caption{Critical values of $h^*$ ($h^*_{\mathrm{c}}$) with $\nu$ degrees of freedom from $n$ i.i.d. random variables of log-normal distribution, i.e.,~$\ln X_k \sim \mathcal{N}\left(0,1\right)$}
    \label{tab:log-normal-table}
    \begin{tabular}{*{11}c}
      \multirow{4}{*}{\diagbox[innerwidth=1.2em,height=3em,linewidth=.5pt,dir=SW,innerleftsep=-6pt,innerrightsep=0pt]{$\left.n\,\middle|\,\nu\right.$}{$h^*_{\mathrm{c}}$}} & \multicolumn{10}{c}{\textbf{Single-tail Signifiance Level}} \\    
       & 0.40 & 0.30 & 0.20 & 0.15 & 0.1 & 0.05 & 0.02 & 0.01 & 0.002 & 0.001 \\
       & \multicolumn{10}{c}{\textbf{Confidence Level}} \\
       & 60\% & 70\% & 80\% & 85\% & 90\% & 95\% & 98\% & 99\% & 99.8\% & 99.9\% \\
      \hline 
    4 \textbar\ 2 & 3.1264 & 4.1849 & 6.0320 & 7.6322 & 10.3673 & 16.6612 & 29.2899 & 43.5578 & 103.5157 & 148.1612 \\
    5 \textbar\ 3 & 2.9564 & 3.7977 & 5.2093 & 6.3863 & 8.3226 & 12.5020 & 20.1394 & 27.9586 & 55.6926 & 73.3729 \\
    6 \textbar\ 4 & 2.9245 & 3.6700 & 4.8979 & 5.9040 & 7.5319 & 10.9565 & 16.9644 & 22.8891 & 42.6792 & 54.5425 \\
    7 \textbar\ 5 & 2.9369 & 3.6275 & 4.7533 & 5.6671 & 7.1320 & 10.1712 & 15.4009 & 20.4584 & 36.8665 & 46.4418 \\
    8 \textbar\ 6 & 2.9662 & 3.6218 & 4.6822 & 5.5380 & 6.8994 & 9.6992 & 14.4542 & 18.9948 & 33.4947 & 41.8604 \\
    9 \textbar\ 7 & 3.0033 & 3.6343 & 4.6491 & 5.4638 & 6.7557 & 9.3933 & 13.8381 & 18.0571 & 31.3896 & 38.9706 \\
    10 \textbar\ 8 & 3.0436 & 3.6568 & 4.6392 & 5.4241 & 6.6648 & 9.1856 & 13.4059 & 17.3831 & 29.8635 & 36.8858 \\
    11 \textbar\ 9 & 3.0857 & 3.6856 & 4.6428 & 5.4055 & 6.6078 & 9.0413 & 13.0932 & 16.8927 & 28.7209 & 35.3703 \\
    12 \textbar\ 10 & 3.1275 & 3.7168 & 4.6542 & 5.3990 & 6.5705 & 8.9340 & 12.8534 & 16.5181 & 27.8447 & 34.1842 \\
    13 \textbar\ 11 & 3.1696 & 3.7504 & 4.6727 & 5.4041 & 6.5510 & 8.8600 & 12.6712 & 16.2142 & 27.1851 & 33.2826 \\
    14 \textbar\ 12 & 3.2106 & 3.7847 & 4.6939 & 5.4140 & 6.5422 & 8.8027 & 12.5315 & 15.9833 & 26.6243 & 32.5399 \\
    15 \textbar\ 13 & 3.2508 & 3.8194 & 4.7179 & 5.4293 & 6.5403 & 8.7657 & 12.4190 & 15.8001 & 26.1627 & 31.9184 \\
    16 \textbar\ 14 & 3.2896 & 3.8534 & 4.7433 & 5.4458 & 6.5436 & 8.7356 & 12.3232 & 15.6475 & 25.8166 & 31.4548 \\
    17 \textbar\ 15 & 3.3287 & 3.8889 & 4.7721 & 5.4683 & 6.5548 & 8.7215 & 12.2630 & 15.5328 & 25.5272 & 31.0195 \\
    18 \textbar\ 16 & 3.3661 & 3.9232 & 4.7997 & 5.4902 & 6.5663 & 8.7089 & 12.2051 & 15.4281 & 25.2384 & 30.6472 \\
    19 \textbar\ 17 & 3.4028 & 3.9573 & 4.8286 & 5.5140 & 6.5819 & 8.7036 & 12.1603 & 15.3405 & 25.0076 & 30.3392 \\
    20 \textbar\ 18 & 3.4379 & 3.9900 & 4.8571 & 5.5381 & 6.5977 & 8.7025 & 12.1267 & 15.2719 & 24.8250 & 30.1103 \\
    21 \textbar\ 19 & 3.4728 & 4.0234 & 4.8872 & 5.5650 & 6.6183 & 8.7095 & 12.1068 & 15.2300 & 24.6515 & 29.8351 \\
    22 \textbar\ 20 & 3.5070 & 4.0563 & 4.9162 & 5.5909 & 6.6387 & 8.7161 & 12.0823 & 15.1718 & 24.4997 & 29.6400 \\
    23 \textbar\ 21 & 3.5397 & 4.0874 & 4.9449 & 5.6171 & 6.6605 & 8.7264 & 12.0737 & 15.1340 & 24.3712 & 29.4295 \\
    24 \textbar\ 22 & 3.5725 & 4.1193 & 4.9743 & 5.6442 & 6.6831 & 8.7367 & 12.0576 & 15.0998 & 24.2505 & 29.2499 \\
    25 \textbar\ 23 & 3.6041 & 4.1499 & 5.0024 & 5.6700 & 6.7050 & 8.7507 & 12.0566 & 15.0753 & 24.1698 & 29.1437 \\
    26 \textbar\ 24 & 3.6352 & 4.1804 & 5.0319 & 5.6981 & 6.7300 & 8.7673 & 12.0555 & 15.0533 & 24.0650 & 28.9897 \\
    27 \textbar\ 25 & 3.6660 & 4.2107 & 5.0606 & 5.7251 & 6.7546 & 8.7827 & 12.0579 & 15.0427 & 24.0238 & 28.9330 \\
    28 \textbar\ 26 & 3.6951 & 4.2392 & 5.0880 & 5.7516 & 6.7778 & 8.8003 & 12.0609 & 15.0306 & 23.9505 & 28.8062 \\
    29 \textbar\ 27 & 3.7249 & 4.2689 & 5.1167 & 5.7785 & 6.8026 & 8.8212 & 12.0693 & 15.0295 & 23.8965 & 28.7377 \\
    30 \textbar\ 28 & 3.7527 & 4.2965 & 5.1435 & 5.8045 & 6.8274 & 8.8391 & 12.0773 & 15.0251 & 23.8500 & 28.6765 \\
    31 \textbar\ 29 & 3.7805 & 4.3239 & 5.1696 & 5.8297 & 6.8497 & 8.8578 & 12.0878 & 15.0288 & 23.8051 & 28.6029 \\
    32 \textbar\ 30 & 3.8084 & 4.3519 & 5.1974 & 5.8569 & 6.8762 & 8.8812 & 12.0996 & 15.0268 & 23.7848 & 28.5123 \\
    42 \textbar\ 40 & 4.0567 & 4.6035 & 5.4488 & 6.1056 & 7.1176 & 9.0983 & 12.2601 & 15.1256 & 23.6447 & 28.2444 \\
    52 \textbar\ 50 & 4.2686 & 4.8231 & 5.6739 & 6.3337 & 7.3458 & 9.3224 & 12.4651 & 15.2999 & 23.6739 & 28.1729 \\
    62 \textbar\ 60 & 4.4537 & 5.0170 & 5.8750 & 6.5390 & 7.5567 & 9.5391 & 12.6809 & 15.5055 & 23.8210 & 28.2636 \\
    72 \textbar\ 70 & 4.6186 & 5.1910 & 6.0570 & 6.7251 & 7.7487 & 9.7368 & 12.8822 & 15.7082 & 24.0183 & 28.4710 \\
    82 \textbar\ 80 & 4.7687 & 5.3506 & 6.2261 & 6.9002 & 7.9315 & 9.9297 & 13.0898 & 15.9189 & 24.2106 & 28.6343 \\
    92 \textbar\ 90 & 4.9059 & 5.4963 & 6.3811 & 7.0599 & 8.0992 & 10.1115 & 13.2781 & 16.1188 & 24.4045 & 28.8423 \\
    102 \textbar\ 100 & 5.0333 & 5.6323 & 6.5265 & 7.2119 & 8.2579 & 10.2794 & 13.4632 & 16.3094 & 24.6392 & 29.0914 \\
    202 \textbar\ 200 & 5.9768 & 6.6440 & 7.6245 & 8.3635 & 9.4816 & 11.6290 & 14.9777 & 17.9518 & 26.5544 & 31.0777 \\
    302 \textbar\ 300 & 6.6190 & 7.3358 & 8.3804 & 9.1624 & 10.3412 & 12.5892 & 16.0815 & 19.1630 & 28.0634 & 32.6891 \\
    402 \textbar\ 400 & 7.1205 & 7.8762 & 8.9721 & 9.7900 & 11.0153 & 13.3512 & 16.9649 & 20.1386 & 29.3208 & 34.1414 \\
    502 \textbar\ 500 & 7.5331 & 8.3210 & 9.4621 & 10.3097 & 11.5770 & 13.9887 & 17.7036 & 20.9639 & 30.2756 & 35.1671 \\
    602 \textbar\ 600 & 7.8879 & 8.7040 & 9.8838 & 10.7584 & 12.0623 & 14.5352 & 18.3614 & 21.7216 & 31.3308 & 36.3717 \\
    702 \textbar\ 700 & 8.2034 & 9.0448 & 10.2576 & 11.1564 & 12.4915 & 15.0068 & 18.8739 & 22.2968 & 32.0343 & 37.0862 \\
    802 \textbar\ 800 & 8.4839 & 9.3480 & 10.5916 & 11.5110 & 12.8786 & 15.4569 & 19.4183 & 22.8922 & 32.8347 & 37.9776 \\
    902 \textbar\ 900 & 8.7387 & 9.6215 & 10.8868 & 11.8239 & 13.2153 & 15.8312 & 19.8412 & 23.3692 & 33.4557 & 38.7925 \\
    1002 \textbar\ 1000 & 8.9750 & 9.8758 & 11.1671 & 12.1220 & 13.5422 & 16.2139 & 20.3033 & 23.8808 & 33.9822 & 39.3243
    \end{tabular}
\end{table}
\pagebreak

\section{Results of $h^*$ analysis}
{
    \footnotesize
    \begin{longtable}{cccccc}
        
        \caption{Outlier analysis results of the pretest and posttest for the loneliness intervention}
        \label{tab1} \\
        \toprule
        \textbf{Item} & \textbf{OutlierPosition} & \textbf{CandidateOutliers} & \textbf{DistFitPVal} & \textbf{$h^*$TestPVal} & \textbf{Inference} \\
        \midrule
        \endfirsthead
        \caption[]{\emph{(Continued)}} \\
        \toprule
        \textbf{Item} & \textbf{OutlierPosition} & \textbf{CandidateOutliers} & \textbf{DistFitPVal} & \textbf{$h^*$TestPVal} & \textbf{Inference} \\
        \midrule
        \endhead
        \midrule
        \endfoot
        \bottomrule
        \endlastfoot
        
        MRel & Max & 28: 6.00 & .4107 & 28: .9626 & Do not reject \\
        MRel & Max & \begin{tabular}[t]{@{}r@{}}12: 6.00\\ 28: 6.00\end{tabular} & .3522 & \begin{tabular}[t]{@{}r@{}}12: .9258\\ 28: .9258\end{tabular} & Do not reject \\
        MRel & Max & \begin{tabular}[t]{@{}r@{}}12: 6.00\\ 28: 6.00\\ 15: 5.75\end{tabular} & .3078 & \begin{tabular}[t]{@{}r@{}}12: .8794\\ 28: .8794\\ 15: .9772\end{tabular} & Do not reject \\
        \addlinespace
        MRel & Min & 36: -1.00 & .5939 & 36: .1234 & Do not reject \\
        MRel & Min & \begin{tabular}[t]{@{}r@{}}36: -1.00\\ 41: -2.00\end{tabular} & .6604 & \begin{tabular}[t]{@{}r@{}}36: .0790\\ 41: .6665\end{tabular} & Do not reject \\
        MRel & Min & \begin{tabular}[t]{@{}r@{}}36: -1.00\\ 39: -2.00\\ 41: -2.00\end{tabular} & .7288 & \begin{tabular}[t]{@{}r@{}}36: .0450\\ 39: .5119\\ 41: .5119\end{tabular} & Do not reject \\
        \addlinespace
        MNov & Max & 10: 5.60 & .744 & 10: .6270 & Do not reject \\
        MNov & Max & \begin{tabular}[t]{@{}r@{}}10: 5.60\\ 20: 5.20\end{tabular} & .7537 & \begin{tabular}[t]{@{}r@{}}10: .5301\\ 20: .9264\end{tabular} & Do not reject \\
        MNov & Max & \begin{tabular}[t]{@{}r@{}}10: 5.60\\ 19: 5.20\\ 20: 5.20\end{tabular} & .7514 & \begin{tabular}[t]{@{}r@{}}10: .4415\\ 19: .8639\\ 20: .8639\end{tabular} & Do not reject \\
        \addlinespace
        MNov & Min & 18: -1.80 & .652 & 18: .4234 & Do not reject \\
        MNov & Min & \begin{tabular}[t]{@{}r@{}}18: -1.80\\ 37: -2.40\end{tabular} & .6232 & \begin{tabular}[t]{@{}r@{}}18: .3413\\ 37: .9368\end{tabular} & Do not reject \\
        MNov & Min & \begin{tabular}[t]{@{}r@{}}18: -1.80\\ 37: -2.40\\ 39: -2.60\end{tabular} & .625 & \begin{tabular}[t]{@{}r@{}}18: .2817\\ 37: .8969\\ 39: .9857\end{tabular} & Do not reject \\
        \addlinespace
        MValB & Max & 9: 5.38 & .8276 & 9: .2120 & Do not reject \\
        MValB & Max & \begin{tabular}[t]{@{}r@{}}9: 5.38\\ 19: 5.13\end{tabular} & .9025 & \begin{tabular}[t]{@{}r@{}}9: .1116\\ 19: .2168\end{tabular} & Do not reject \\
        MValB & Max & \begin{tabular}[t]{@{}r@{}}9: 5.38\\ 17: 5.13\\ 19: 5.13\end{tabular} & .8985 & \begin{tabular}[t]{@{}r@{}}9: .0441\\ 17: .0967\\ 19: .0967\end{tabular} & Do not reject \\
        \addlinespace
        MValB & Min & 16: -1.00 & .6427 & 16: .9787 & Do not reject \\
        MValB & Min & \begin{tabular}[t]{@{}r@{}}16: -1.00\\ 25: -1.13\end{tabular} & .5966 & \begin{tabular}[t]{@{}r@{}}16: .9592\\ 25: .9876\end{tabular} & Do not reject \\
        MValB & Min & \begin{tabular}[t]{@{}r@{}}16: -1.00\\ 2: -1.13\\ 25: -1.13\end{tabular} & .5399 & \begin{tabular}[t]{@{}r@{}}16: .9258\\ 2: .9725\\ 25: .9725\end{tabular} & Do not reject \\
        \addlinespace
        MValA & Max & \begin{tabular}[t]{@{}r@{}}17: 5.75\end{tabular} & .7888 & \begin{tabular}[t]{@{}r@{}}17: .7514\end{tabular} & Do not reject \\
        MValA & Max & \begin{tabular}[t]{@{}r@{}}17: 5.75\\ 9: 5.38\end{tabular} & .7492 & \begin{tabular}[t]{@{}r@{}}17: .6734\\ 9: .9772\end{tabular} & Do not reject \\
        MValA & Max & \begin{tabular}[t]{@{}r@{}}17: 5.75\\ 9: 5.38\\ 19: 5.25\end{tabular} & 0.721 & \begin{tabular}[t]{@{}r@{}}17: .6082\\ 9: .9578\\ 19: .9923\end{tabular} & Do not reject \\
        \addlinespace
        MValA & Min & 39: -1.75 & .9333 & 39: .0417 & Reject \\
        MValA & Min & \begin{tabular}[t]{@{}r@{}}39: -1.75\\ 41: -2.63\end{tabular} & .9302 & \begin{tabular}[t]{@{}r@{}}39: .0227\\ 41: .5087\end{tabular} & Do not reject \\
        MValA & Min & \begin{tabular}[t]{@{}r@{}}39: -1.75\\ 41: -2.63\\ 27: -2.75\end{tabular} & .9008 & \begin{tabular}[t]{@{}r@{}}39: .0117\\ 41: .3647\\ 27: .5149\end{tabular} & Do not reject \\
        \addlinespace
        MValAi & Max & 17: 5.50 & .1105 & 17: .9465 & Do not reject \\
        MValAi & Max & \begin{tabular}[t]{@{}r@{}}17: 5.50\\ 9: 5.25\end{tabular} & .0938 & \begin{tabular}[t]{@{}r@{}}17: .9121\\ 9: .9897\end{tabular} & Do not reject \\
        MValAi & Max & \begin{tabular}[t]{@{}r@{}}17: 5.50\\ 9: 5.25\\ 16: 5.00\end{tabular} & .0834 & \begin{tabular}[t]{@{}r@{}}17: .8872\\ 9: .9813\\ 16: .9994\end{tabular} & Do not reject \\
        \addlinespace
        MValAi & Min & 33: -1.00 & .1714 & 33: .1276 & Do not reject \\
        MValAi & Min & \begin{tabular}[t]{@{}r@{}}33: -1.00\\ 26: -1.50\end{tabular} & .2079 & \begin{tabular}[t]{@{}r@{}}33: .0653\\ 26: .2220\end{tabular} & Do not reject \\
        MValAi & Min & \begin{tabular}[t]{@{}r@{}}33: -1.00\\ 26: -1.50\\ 39: -1.75\end{tabular} & .2486 & \begin{tabular}[t]{@{}r@{}}33: .0318\\ 26: .1535\\ 39: .3069\end{tabular} & Do not reject \\
        \addlinespace
        MValAc & Max & 39: 5.25 & .7146 & 39: .6291 & Reject \\
        MValAc & Max & \begin{tabular}[t]{@{}r@{}}39: 5.25\\ 27: 4.25\end{tabular} & .6441 & \begin{tabular}[t]{@{}r@{}}39: .0182\\ 27: .9873\end{tabular} & Do not reject \\
        MValAc & Max & \begin{tabular}[t]{@{}r@{}}39: 5.25\\ 27: 4.25\\ 41: 4.00\end{tabular} & .5709 & \begin{tabular}[t]{@{}r@{}}39: .0072\\ 27: .2641\\ 41: .5282\end{tabular} & Do not reject \\
        \addlinespace
        MValAc & Min & \begin{tabular}[t]{@{}r@{}}38: -1.00\end{tabular} & .7323 & \begin{tabular}[t]{@{}r@{}}38: .9882\end{tabular} & Do not reject \\
        MValAc & Min & \begin{tabular}[t]{@{}r@{}}38: -1.00\\ 38: -1.00\end{tabular} & .7105 & \begin{tabular}[t]{@{}r@{}}33: .9739\\ 38: .9739\end{tabular} & Do not reject \\
        MValAc & Min & \begin{tabular}[t]{@{}r@{}}19: -1.00\\ 33: -1.00\\ 38: -1.00\end{tabular} & .6745 & \begin{tabular}[t]{@{}r@{}}19: .9459\\ 33: .9459\\ 38: .9459\end{tabular} & Do not reject \\
        \addlinespace
        MValBc & Max & 25: 6.00 & .3653 & 25: .9655 & Do not reject \\
        MValBc & Max & \begin{tabular}[t]{@{}r@{}}16: 6.00\\ 25: 6.00\end{tabular} & 0.311 & \begin{tabular}[t]{@{}r@{}}16: .9310\\ 25: .9310\end{tabular} & Do not reject \\
        MValBc & Max & \begin{tabular}[t]{@{}r@{}}2: 6.00\\ 16: 6.00\\ 25: 6.00\end{tabular} & .2545 & \begin{tabular}[t]{@{}r@{}}2: .8714\\ 16: .8714\\ 25: .8714\end{tabular} & Do not reject \\
        \addlinespace
        MValBc & Min & 19: -1.00 & .4944 & 19: .2455 & Do not reject \\
        MValBc & Min & \begin{tabular}[t]{@{}r@{}}19: -1.00\\ 9: -1.75\end{tabular} & 0.559 & \begin{tabular}[t]{@{}r@{}}19: .1673\\ 9: .6666\end{tabular} & Do not reject \\
        MValBc & Min & \begin{tabular}[t]{@{}r@{}}19: -1.00\\ 9: -1.75\\ 33: -2.00\end{tabular} & .6213 & \begin{tabular}[t]{@{}r@{}}19: .1140\\ 9: .5434\\ 33: .7613\end{tabular} & Do not reject \\
        \addlinespace
        MValBi & Max & 9: 5.50 & .5106 & 9: .0615 & Do not reject \\
        MValBi & Max & \begin{tabular}[t]{@{}r@{}}9: 5.50\\ 17: 5.25\end{tabular} & .4262 & \begin{tabular}[t]{@{}r@{}}9: .0181\\ 17: .0403\end{tabular} & Reject \\
        MValBi & Max & \begin{tabular}[t]{@{}r@{}}9: 5.50\\ 17: 5.25\\ 19: 4.25\end{tabular} & .3596 & \begin{tabular}[t]{@{}r@{}}9: .0086\\ 17: .0204\\ 19: .4142\end{tabular} & Do not reject \\
        \addlinespace
        MValBi & Min & 41: -1.00 & .4873 & 41: .9998 & Do not reject \\
        MValBi & Min & \begin{tabular}[t]{@{}r@{}}18: -1.00\\ 41: -1.00\end{tabular} & .4897 & \begin{tabular}[t]{@{}r@{}}18: .9995\\ 41: .9995\end{tabular} & Do not reject \\
        MValBi & Min & \begin{tabular}[t]{@{}r@{}}18: -1.00\\ 18: -1.00\\ 41: -1.00\end{tabular} & .4821 & \begin{tabular}[t]{@{}r@{}}18: .9987\\ 18: .9987\\ 41: .9987\end{tabular} & Do not reject \\
        \addlinespace
        DVal & Max & 16: 4.00 & .5525 & 16: .3846 & Do not reject \\
        DVal & Max & \begin{tabular}[t]{@{}r@{}}16: 4.00\\ 11: 3.50\end{tabular} & .5749 & \begin{tabular}[t]{@{}r@{}}16: .2764\\ 11: .6947\end{tabular} & Do not reject \\
        DVal & Max & \begin{tabular}[t]{@{}r@{}}16: 4.00\\ 11: 3.50\\ 13: 3.38\end{tabular} & .5952 & \begin{tabular}[t]{@{}r@{}}16: .1930\\ 11: .5619\\ 13: .6778\end{tabular} & Do not reject \\
        \addlinespace
        DVal & Min & 5: 0.25 & .5222 & 5: .9884 & Do not reject \\
        DVal & Min & \begin{tabular}[t]{@{}r@{}}5: 0.25\\ 39: 0.00\end{tabular} & .5394 & \begin{tabular}[t]{@{}r@{}}5: .9787\\ 39: .9990\end{tabular} & Do not reject \\
        DVal & Min & \begin{tabular}[t]{@{}r@{}}5: 0.25\\ 9: 0.00\\ 39: 0.00\end{tabular} & .5459 & \begin{tabular}[t]{@{}r@{}}5: .9622\\ 9: .9972\\ 39: .9972\end{tabular} & Do not reject \\
        \addlinespace
        DValP & Max & 16: 4.00 & .8846 & 16: .5871 & Do not reject \\
        DValP & Max & \begin{tabular}[t]{@{}r@{}}16: 4.00\\ 13: 3.50\end{tabular} & .8843 & \begin{tabular}[t]{@{}r@{}}16: .4792\\ 13: .8573\end{tabular} & Do not reject \\
        DValP & Max & \begin{tabular}[t]{@{}r@{}}16: 4.00\\ 13: 3.50\\ 12: 3.25\end{tabular} & 0.892 & \begin{tabular}[t]{@{}r@{}}16: .3892\\ 13: .7808\\ 12: .9285\end{tabular} & Do not reject \\
        \addlinespace
        DValP & Min & 26: 1.00 & .8774 & 26: .8353 & Do not reject \\
        DValP & Min & \begin{tabular}[t]{@{}r@{}}26: 1.00\\ 34: 0.75\end{tabular} & .8561 & \begin{tabular}[t]{@{}r@{}}26: .7512\\ 34: .9066\end{tabular} & Do not reject \\
        DValP & Min & \begin{tabular}[t]{@{}r@{}}26: 1.00\\ 34: 0.75\\ 39: 0.50\end{tabular} & .8454 & \begin{tabular}[t]{@{}r@{}}26: .6667\\ 34: .8499\\ 39: .9615\end{tabular} & Do not reject \\
        \addlinespace
        DValN & Max & 16: 4.00 & .7014 & 16: .5128 & Do not reject \\
        DValN & Max & \begin{tabular}[t]{@{}r@{}}1: 4.00\\ 16: 4.00\end{tabular} & .7529 & \begin{tabular}[t]{@{}r@{}}1: .3482\\ 16: .3482\end{tabular} & Do not reject \\
        DValN & Max & \begin{tabular}[t]{@{}r@{}}1: 4.00\\ 16: 4.00\\ 11: 3.75\end{tabular} & .7867 & \begin{tabular}[t]{@{}r@{}}1: .2211\\ 16: .2211\\ 11: .3913\end{tabular} & Do not reject \\
        \addlinespace
        DValN & Min & 27: 0.25 & 0.595 & 27: .9734 & Do not reject \\
        DValN & Min & \begin{tabular}[t]{@{}r@{}}27: 0.25\\ 40: 0.00\end{tabular} & .5777 & \begin{tabular}[t]{@{}r@{}}27: .9534\\ 40: .9957\end{tabular} & Do not reject \\
        DValN & Min & \begin{tabular}[t]{@{}r@{}}27: 0.25\\ 19: 0.00\\ 40: 0.00\end{tabular} & .5471 & \begin{tabular}[t]{@{}r@{}}27: .9216\\ 19: .9898\\ 40: .9898\end{tabular} & Do not reject \\
    \end{longtable}
}
\clearpage

\section{Analysis of paired-samples test on $h^*$}
The example analysis in Section~\ref{sec:paired-test} takes the following raw scores:
\begin{table}[h]
    \centering
    \caption{Generated raw loneliness scores for 180 participants before ($x_{\text{pre}}$) and after interventions ($x_{\text{post}}$)}
    \label{tab:paired-test-raw}
\begin{tabular}{ccc|ccc|ccc|ccc}
 \# & $x_{\text{pre}}$ & $x_{\text{post}}$ & \# & $x_{\text{pre}}$ & $x_{\text{post}}$ & \# & $x_{\text{pre}}$ & $x_{\text{post}}$ & \# & $x_{\text{pre}}$ & $x_{\text{post}}$ \\
 \hline
 1 & 1.34 & 1.51 & 46 & 1.34 & 1.45 & 91 & 1.31 & 1.24 & 136 & 1.32 & 1.18 \\
 2 & 1.57 & 1.38 & 47 & 1.19 & 1.07 & 92 & 1.20 & 1.13 & 137 & 1.25 & 1.33 \\
 3 & 1.41 & 1.60 & 48 & 1.67 & 1.57 & 93 & 1.46 & 1.57 & 138 & 1.38 & 1.25 \\
 4 & 1.31 & 1.19 & 49 & 0.88 & 0.77 & 94 & 1.14 & 1.19 & 139 & 1.12 & 1.39 \\
 5 & 1.17 & 1.25 & 50 & 1.47 & 1.29 & 95 & 1.52 & 1.34 & 140 & 1.53 & 1.12 \\
 6 & 1.44 & 1.23 & 51 & 0.98 & 0.71 & 96 & 1.10 & 1.24 & 141 & 1.09 & 1.57 \\
 7 & 1.21 & 1.09 & 52 & 1.58 & 1.74 & 97 & 1.28 & 1.43 & 142 & 1.28 & 1.08 \\
 8 & 1.60 & 1.87 & 53 & 1.21 & 1.06 & 98 & 1.48 & 1.13 & 143 & 1.50 & 1.30 \\
 9 & 1.69 & 1.54 & 54 & 1.09 & 1.27 & 99 & 1.17 & 1.56 & 144 & 1.19 & 1.51 \\
 10 & 1.22 & 1.13 & 55 & 1.36 & 1.54 & 100 & 1.33 & 1.08 & 145 & 1.35 & 1.17 \\
 11 & 1.32 & 1.23 & 56 & 1.55 & 1.36 & 101 & 1.25 & 1.29 & 146 & 1.23 & 1.37 \\
 12 & 1.54 & 1.73 & 57 & 1.41 & 1.24 & 102 & 1.39 & 1.51 & 147 & 1.40 & 1.22 \\
 13 & 1.36 & 1.19 & 58 & 1.13 & 1.23 & 103 & 1.12 & 1.16 & 148 & 1.16 & 1.41 \\
 14 & 1.19 & 1.33 & 59 & 2.76 & 2.14 & 104 & 1.54 & 1.36 & 149 & 1.54 & 1.14 \\
 15 & 1.56 & 1.44 & 60 & 0.82 & 0.87 & 105 & 1.09 & 1.21 & 150 & 1.11 & 1.53 \\
 16 & 1.25 & 1.41 & 61 & 1.25 & 1.37 & 106 & 1.26 & 1.43 & 151 & 1.31 & 1.12 \\
 17 & 1.20 & 1.06 & 62 & 1.55 & 1.36 & 107 & 1.50 & 1.14 & 152 & 1.20 & 1.29 \\
 18 & 1.50 & 1.32 & 63 & 0.74 & 0.78 & 108 & 1.19 & 1.55 & 153 & 1.46 & 1.47 \\
 19 & 1.63 & 1.45 & 64 & 1.09 & 1.27 & 109 & 1.37 & 1.10 & 154 & 1.15 & 1.19 \\
 20 & 1.23 & 1.42 & 65 & 1.53 & 1.34 & 110 & 1.22 & 1.27 & 155 & 1.52 & 1.34 \\
 21 & 1.38 & 1.24 & 66 & 1.34 & 1.45 & 111 & 1.41 & 1.49 & 156 & 1.10 & 1.24 \\
 22 & 1.13 & 1.05 & 67 & 1.19 & 1.07 & 112 & 1.15 & 1.17 & 157 & 1.26 & 1.43 \\
 23 & 1.47 & 1.61 & 68 & 2.95 & 2.22 & 113 & 1.53 & 1.32 & 158 & 2.95 & 2.10 \\
 24 & 1.09 & 1.19 & 69 & 0.81 & 0.77 & 114 & 1.11 & 1.25 & 159 & 1.18 & 1.07 \\
 25 & 1.53 & 1.41 & 70 & 1.47 & 1.29 & 115 & 1.30 & 1.38 & 160 & 1.33 & 1.23 \\
 26 & 2.34 & 2.25 & 71 & 1.12 & 1.29 & 116 & 1.47 & 1.12 & 161 & 1.25 & 1.36 \\
 27 & 1.19 & 1.00 & 72 & 1.38 & 1.19 & 117 & 1.18 & 1.53 & 162 & 1.39 & 1.51 \\
 28 & 1.67 & 1.77 & 73 & 1.06 & 1.23 & 118 & 1.34 & 1.09 & 163 & 1.13 & 1.62 \\
 29 & 1.00 & 0.73 & 74 & 1.24 & 1.43 & 119 & 1.24 & 1.28 & 164 & 1.57 & 1.12 \\
 30 & 1.47 & 1.29 & 75 & 1.15 & 1.22 & 120 & 1.42 & 1.50 & 165 & 1.08 & 1.79 \\
 31 & 0.94 & 0.84 & 76 & 1.43 & 1.16 & 121 & 1.13 & 1.19 & 166 & 1.30 & 1.44 \\
 32 & 1.58 & 1.39 & 77 & 1.29 & 1.13 & 122 & 1.56 & 1.35 & 167 & 1.51 & 1.28 \\
 33 & 1.21 & 1.09 & 78 & 1.51 & 1.70 & 123 & 1.08 & 1.23 & 168 & 1.17 & 1.85 \\
 34 & 1.09 & 1.25 & 79 & 1.11 & 1.32 & 124 & 1.29 & 1.40 & 169 & 1.37 & 1.12 \\
 35 & 1.36 & 1.54 & 80 & 1.60 & 1.44 & 125 & 1.51 & 1.16 & 170 & 1.22 & 1.17 \\
 36 & 1.55 & 1.36 & 81 & 1.08 & 1.23 & 126 & 1.16 & 1.54 & 171 & 1.41 & 1.73 \\
 37 & 1.41 & 1.24 & 82 & 1.27 & 1.45 & 127 & 1.36 & 1.11 & 172 & 1.14 & 1.04 \\
 38 & 1.13 & 1.23 & 83 & 1.49 & 1.33 & 128 & 1.21 & 1.31 & 173 & 3.22 & 2.37 \\
 39 & 1.62 & 1.81 & 84 & 1.18 & 1.09 & 129 & 1.43 & 1.20 & 174 & 1.12 & 1.32 \\
 40 & 0.73 & 0.87 & 85 & 1.35 & 1.41 & 130 & 1.14 & 1.46 & 175 & 1.29 & 1.68 \\
 41 & 1.25 & 1.37 & 86 & 1.23 & 1.36 & 131 & 1.55 & 1.15 & 176 & 1.47 & 1.09 \\
 42 & 1.55 & 1.36 & 87 & 1.40 & 1.23 & 132 & 1.10 & 1.52 & 177 & 2.90 & 2.14 \\
 43 & 1.01 & 0.78 & 88 & 1.16 & 1.09 & 133 & 1.27 & 1.10 & 178 & 1.34 & 1.55 \\
 44 & 1.09 & 1.27 & 89 & 1.55 & 1.36 & 134 & 1.49 & 1.26 & 179 & 1.24 & 1.21 \\
 45 & 1.53 & 1.34 & 90 & 1.07 & 1.26 & 135 & 1.17 & 1.48 & 180 & 1.43 & 1.39 \\
\end{tabular}
\end{table}

Suspecting a log-normal distribution, log-scores were used for evaluating the $h^*$ values. The pretest results were as follows:

{
    \footnotesize
    \begin{longtable}{cccccc}
        
        \caption{Results of the outlier analysis, showing candidate outliers, distribution fit $p$-value, the $h^*$-test $p$-value, and the conclusion of outlier inference.}
        \label{tab:paired-test-result} \\
        \toprule
        \textbf{Item} & \textbf{OutlierPosition} & \textbf{CandidateOutliers} & \textbf{DistFitPVal} & \textbf{$h^*$TestPVal} & \textbf{Inference} \\
        \midrule
        \endfirsthead
        \caption[]{\emph{(Continued)}} \\
        \toprule
        \textbf{Item} & \textbf{OutlierPosition} & \textbf{CandidateOutliers} & \textbf{DistFitPVal} & \textbf{$h^*$TestPVal} & \textbf{Inference} \\
        \midrule
        \endhead
        \midrule
        \endfoot
        \bottomrule
        \endlastfoot
Pretest & Max & 173: 1.17 & .0827 & 173: .0010 & Reject \\
Pretest & Max & \begin{tabular}[r]{@{}l@{}}173: 1.17\\ 68: 1.08\end{tabular} & .1569 & \begin{tabular}[r]{@{}l@{}}173: .0004\\ 68: .0025\end{tabular} & Reject \\
Pretest & Max & \begin{tabular}[r]{@{}l@{}}173: 1.17\\ 68: 1.08\\ 158: 1.08\end{tabular} & .2827 & \begin{tabular}[r]{@{}l@{}}173: .0001\\ 68: .0009\\ 158: .0009\end{tabular} & Reject \\
Pretest & Max & \begin{tabular}[r]{@{}l@{}}173: 1.17\\ 68: 1.08\\ 158: 1.08\\ 177: 1.06\end{tabular} & .4293 & \begin{tabular}[r]{@{}l@{}}173: .0000\\ 68: .0002\\ 158: .0002\\ 177: .0004\end{tabular} & Reject \\
Pretest & Max & \begin{tabular}[r]{@{}l@{}}173: 1.17\\ 68: 1.08\\ 158: 1.08\\ 177: 1.06\\ 59: 1.02\end{tabular} & .4762 & \begin{tabular}[r]{@{}l@{}}173: .0000\\ 68: .0001\\ 158: .0001\\ 177: .0001\\ 59: .0004\end{tabular} & Reject \\
Pretest & Max & \begin{tabular}[r]{@{}l@{}}173: 1.17\\ 68: 1.08\\ 158: 1.08\\ 177: 1.06\\ 59: 1.02\\ 26: 0.85\end{tabular} & .4132 & \begin{tabular}[r]{@{}l@{}}173: .0000\\ 68: .0000\\ 158: .0000\\ 177: .0000\\ 59: .0001\\ 26: .0123\end{tabular} & Reject \\
Pretest & Max & \begin{tabular}[r]{@{}l@{}}173: 1.17\\ 68: 1.08\\ 158: 1.08\\ 177: 1.06\\ 59: 1.02\\ 26: 0.85\\ 9: 0.52\end{tabular} & .4009 & \begin{tabular}[r]{@{}l@{}}173: .0000\\ 68: .0000\\ 158: .0000\\ 177: .0000\\ 59: .0001\\ 26: .0108\\ 9: .9999\end{tabular} & Do not reject \\
Posttest & Max & 173: 0.86 & .1989 & 173: .1827 & Do not reject \\
Posttest & Max & \begin{tabular}[r]{@{}l@{}}173: 0.86\\ 26: 0.81\end{tabular} & .2312 & \begin{tabular}[r]{@{}l@{}}173: .1420\\ 26: .3093\end{tabular} & Do not reject \\
Posttest & Max & \begin{tabular}[r]{@{}l@{}}173: 0.86\\ 26: 0.81\\ 68: 0.80\end{tabular} & .2613 & \begin{tabular}[r]{@{}l@{}}173: .1089\\ 26: .2498\\ 68: .3148\end{tabular} & Do not reject \\
Posttest & Max & \begin{tabular}[r]{@{}l@{}}173: 0.86\\ 26: 0.81\\ 68: 0.80\\ 59: 0.76\end{tabular} & .2826 & \begin{tabular}[r]{@{}l@{}}173: .0838\\ 26: .2016\\ 68: .2584\\ 59: .4093\end{tabular} & Do not reject \\
Posttest & Max & \begin{tabular}[r]{@{}l@{}}173: 0.86\\ 26: 0.81\\ 68: 0.80\\ 177: 0.76\\ 59: 0.76\end{tabular} & .3001 & \begin{tabular}[r]{@{}l@{}}173: .0627\\ 26: .1584\\ 68: .2064\\ 177: .3398\\ 59: .3398\end{tabular} & Do not reject \\
Posttest & Max & \begin{tabular}[r]{@{}l@{}}173: 0.86\\ 26: 0.81\\ 68: 0.80\\ 177: 0.76\\ 59: 0.76\\ 158: 0.74\end{tabular} & .3069 & \begin{tabular}[r]{@{}l@{}}173: .0467\\ 26: .1236\\ 68: .1636\\ 177: .2792\\ 59: .2792\\ 158: .3873\end{tabular} & Do not reject \\
Posttest & Min & 51: 0.35 & .1947 & 51: .1499 & Do not reject \\
Posttest & Min & \begin{tabular}[r]{@{}l@{}}51: 0.35\\ 29: 0.31\end{tabular} & .2237 & \begin{tabular}[r]{@{}l@{}}51: .1113\\ 29: .1835\end{tabular} & Do not reject \\
Posttest & Min & \begin{tabular}[r]{@{}l@{}}51: 0.35\\ 29: 0.31\\ 69: 0.26\end{tabular} & .2386 & \begin{tabular}[r]{@{}l@{}}51: .0844\\ 29: .1432\\ 69: .3260\end{tabular} & Do not reject \\
Posttest & Min & \begin{tabular}[r]{@{}l@{}}51: 0.35\\ 29: 0.31\\ 49: 0.26\\ 69: 0.26\end{tabular} & .2473 & \begin{tabular}[r]{@{}l@{}}51: .0619\\ 29: .1083\\ 49: .2616\\ 69: .2616\end{tabular} & Do not reject \\
Posttest & Min & \begin{tabular}[r]{@{}l@{}}51: 0.35\\ 29: 0.31\\ 49: 0.26\\ 69: 0.26\\ 63: 0.25\end{tabular} & .2453 & \begin{tabular}[r]{@{}l@{}}51: .0446\\ 29: .0802\\ 49: .2055\\ 69: .2055\\ 63: .2527\end{tabular} & Do not reject \\
Posttest & Min & \begin{tabular}[r]{@{}l@{}}51: 0.35\\ 29: 0.31\\ 49: 0.26\\ 69: 0.26\\ 43: 0.25\\ 63: 0.25\end{tabular} & .2329 & \begin{tabular}[r]{@{}l@{}}51: .0309\\ 29: .0574\\ 49: .1559\\ 69: .1559\\ 43: .1949\\ 63: .1949\end{tabular} & Do not reject
\end{longtable}
}
The largest set of outliers was selected in the pretest, recognising six outliers, as tabulated in Table~\ref{tab:paired-h}. No outliers were identified in the posttest. The rest of the analysis is delivered in Section~\ref{sec:paired-test}.

\end{appendix}

\begin{acks}[Acknowledgments]
This study is funded by the Research Grants Council (project code: T43 518/24-N) under the University Grants Committee, Hong Kong Special Administrative Region Government. The authors would like to thank Rain T. C. Au-yeung for assisting with the typesetting of the manuscript.
\end{acks}



\bibliographystyle{imsart-number} 
\bibliography{bibliography}       


\end{document}